\documentclass[%
 aip,
 amsmath,amssymb,
 reprint,%
]{revtex4-1}

\usepackage{graphicx}
\usepackage{dcolumn}
\usepackage{bm}

\usepackage[utf8]{inputenc}
\usepackage[T1]{fontenc}
\usepackage{mathptmx}
\usepackage{etoolbox}
\usepackage[makeroom]{cancel}
\usepackage{multirow}
\usepackage[table]{xcolor}
\usepackage[textsize=scriptsize]{todonotes}
\usepackage{float} 
\usepackage{diagbox}
\usepackage{comment}
\usepackage{tikz}
\usepackage[normalem]{ulem}

\makeatletter
\def\@email#1#2{%
 \endgroup
 \patchcmd{\titleblock@produce}
  {\frontmatter@RRAPformat}
  {\frontmatter@RRAPformat{\produce@RRAP{*#1\href{mailto:#2}{#2}}}\frontmatter@RRAPformat}
  {}{}
}%
\makeatother
\begin{document}

\preprint{AIP/123-QED}

\title[]{A Data-Driven Approach for Predicting Hydrodynamic Forces on Spherical Particles Using Volume Fraction Representations}
\author{A. Metelkin}%
\altaffiliation[Also at ]{Leichtwei\ss-Institute for Hydraulic Engineering and Water Resources, TU Braunschweig, 38106 Braunschweig, Germany}
\affiliation{
Institute of Urban and Industrial Water Management, Technische Universität Dresden,Bergstraße 66, 01069, Dresden, Germany}

\author{S.J. Jacob}
\altaffiliation[Also at ]{Leichtwei\ss-Institute for Hydraulic Engineering and Water Resources, TU Braunschweig, 38106 Braunschweig, Germany}
\affiliation{Chair for System Simulation, Friedrich–Alexander–Universität Erlangen–Nürnberg, Cauerstraße 11,
91058 Erlangen, Germany}

\author{B. Vowinckel}%
\altaffiliation[Also at ]{Leichtwei\ss-Institute for Hydraulic Engineering and Water Resources, TU Braunschweig, 38106 Braunschweig, Germany}
\affiliation{
Institute of Urban and Industrial Water Management, Technische Universität Dresden,Bergstraße 66, 01069, Dresden, Germany}

\email{alexander.metelkin@tu-dresden.de}

\date{\today}

\begin{abstract}
Particle-laden flows are simulated at various scales using numerical techniques that range from particle-resolved Direct Numerical Simulations (pr-DNS) for small-scale systems to Lagrange point-particle methods for laboratory-scale problems, and Euler-Euler approaches for larger-scale applications. Recent research has been particularly focused on the development of both physics-based and data-driven closures to enhance the accuracy of the Lagrangian point-particle approach by leveraging highly resolved data from pr-DNS. In this study, a data-driven methodology is presented for the prediction of hydrodynamic forces acting on spherical particles immersed in an ambient flow field, where neighboring particle information is represented by volume fractions. The volume fractions are computed on an auxiliary grid with cell sizes on the order of the particle diameter. The volume fraction values in the vicinity of each particle are used as input features for the data-driven model to predict the corresponding hydrodynamic forces and moments. The training data was generated by a series of pr-DNS of flow through arrays of randomly distributed, fixed-position particles at various Reynolds numbers and particle volume fractions. The data-driven model is built using Fully Connected Neural Networks (FCNN). Improved prediction accuracy of hydrodynamic forces and torques is demonstrated in comparison to FCNN models that rely on direct particle position inputs. In addition, the proposed volume-fraction-based approach exhibits greater flexibility than previously introduced models by accommodating systems with particles of different sizes and shapes.

\vspace{1em}
\noindent\textbf{The following article has been accepted by Physics of Fluids. After it is published, it will be found at \href{https://doi.org/10.1063/5.0279971}{https://doi.org/10.1063/5.0279971}}

\end{abstract}

\maketitle

\newcommand{\PPB}{PPB }
\newcommand{\LVFB}{LVFB }
\newcommand{\PPBlong}{Particle Position-Based }
\newcommand{\LVFBlong}{Local Volume Fraction-Based }

\newcommand{\review}[1]{\textcolor{black}{#1}}

\section{Introduction}
Suspension flows play a crucial role in a wide range of industrial and environmental processes. A prominent example is sediment transport in aqueous systems, which is essential for engineering and environmental applications. In addition, suspension flows are present in phenomena such as dust storms, volcanic ash dispersal, pneumatic conveying, and spray drying, highlighting the broad applicability of particle-laden flow models across various spatial scales. Simulating these flows presents significant challenges due to the wide range of spatial and temporal scales involved. For example, river lengths can reach up to $\sim100$ km, while individual sand grains measure only millimeters. This disparity in scales requires the use of closures to enhance the precision of macroscale models by integrating microscale data \cite{vowinckel2021}, which requires computational approaches to be adapted to the specific scales of suspended particle systems.

Particle-resolved Direct Numerical Simulation (pr-DNS) is a microscale approach and the most accurate method for simulating suspension flows, as it captures all temporal and spatial scales of both the fluid and solid phases \cite{vowinckel2021}. However, pr-DNS requires substantial computational resources because of the detailed geometric representation of each sediment grain, which usually results in a mesh cell size of $\Delta h \propto D_p / 10$, where $D_p$ denotes the diameter of the particle \cite{uhlmann2005immersed}. This approach is generally restricted to resolving domains with lengths of up to approximately $\sim 10^3 D_p$, limiting the maximum spatial scale to around $\sim 10^0$ meters, which is insufficient for most applications \cite{subramaniam2020multiphase,subramaniam2022modeling}. For intermediate scales, the point-particle simulation method (mesoscale approach) provides a suitable alternative \cite{m2016point}. The method represents particles as discrete entities, capturing their interactions without explicitly resolving the fluid flow around them, offering a balanced trade-off between accuracy and computational efficiency for scales of approximately $\sim 10^{0}-10^{1}$ meters. Understanding the transition between the microscale and the mesoscale is a central focus of the present study.

In the point-particle approach, the surface of the particle is not resolved by the numerical grid. Therefore, a direct calculation of the hydrodynamic force is not possible, and the hydrodynamic forces must be modeled. This force modeling is a primary closure between microscale and mesoscale approaches. Theoretical work that estimates the drag force on particles is mostly limited to the dilute regime of particles in the Eulerian field and mostly to Reynolds numbers ($Re$) less than one \cite{tenneti2011}. 
The main coupling approach comes from the derivation of the drag force that acts on a sphere in flows with low Reynolds numbers proposed by \citet{stokes1901mathematical}. \citet{hasimoto1959periodic} put forward an analytical expression for the drag force on a dilute array of spherical particles in Stokes flow, and \citet{sangani1982slow} expanded this approach by varying the particle volume fractions ($\phi$). While such analytical solutions are useful for simplified and idealized configurations, they are not directly applicable to random particle arrangements, especially in inertial regimes ($Re > 1$). To address the complexities introduced by random particle distributions and higher Reynolds numbers, pr-DNS have become a powerful tool. These simulations resolve the fluid–particle interactions at the scale of individual particles and provide detailed data on the hydrodynamic forces they experience. Early studies employing pr-DNS to quantify average drag forces laid the groundwork for data-driven model development. \citet{HILL_KOCH_LADD_2001} provided a comprehensive summary of these early efforts. Building on this foundation, subsequent studies \cite{beetstra2007drag, tenneti2011, bogner2015, yali2015new,van2024microstructure} have extended the parameter space to include a broader range of $Re$ and $\phi$, improving the accuracy of drag correlations in moderately dense and inertial regimes.

To account for the influence of the relative positions of neighboring particles on the hydrodynamic forces in the mesoscale approach, \citet{akiki2017pairwise} proposed the Pairwise Interaction Extended Point-particle (PIEP) model. This model utilizes pairwise interaction assumptions and incorporates the perturbation effects of the flow generated by neighboring particles. The PIEP model was assessed using randomly arranged fixed particles for volume fractions $\phi \in \{0.1, 0.21\}$ and demonstrated a higher correlation compared to models based solely on volume fraction. Subsequently, \citet{moore2019} enhanced the accuracy of the PIEP model introduced by \citet{akiki2017pairwise}. The improved PIEP model showed a significant improvement in lift force prediction for $\phi \in \{0.1, 0.2\}$, however, this improvement declined as volume fractions increased. To address this limitation, the authors developed a hybrid model that combines the PIEP model with a statistical-based approach. The hybrid model showed improved accuracy in terms of the coefficient of determination ($R^2$) scores for both lift and drag components of the force, particularly for systems with higher volume fractions. \citet{balachandar2020} conducted a comparative analysis of a similar hybrid model introduced by \citet{moore2019} against a fundamentally different approach that does not directly account for flow modifications. This alternative approach, presented by \citet{he2019supervised}, is purely data-driven and employs a Fully Connected Neural Network (FCNN) to predict the forces on the particles without reconstructing the flow field. The comparison revealed that the latter approach underperformed relative to the hybrid model. Independently, \citet{seyed2020} introduced the microstructure-informed probability-driven point-particle (MPP) model, which utilizes statistical information extracted from pr-DNS to construct probability distribution functions for forces and torques. These functions can be used to predict the perturbation components of the corresponding forces and torques. However, the authors noted that the construction of a reliable data set for a regression-based model requires millions of samples (particles) for each combination of Reynolds number \(Re\) and volume fraction \(\phi\). Subsequently, \citet{seyed2021} also developed a Physics-Informed Neural Network (PINN) \cite{seyed2021}, which similarly to the PIEP model, utilizes the pairwise particle interaction assumption that employs the treatment of each individual neighbor via a separate NN block. This approach further enhanced the accuracy of the hydrodynamic force and torque predictions for the particles. Building on this work, \citet{cheng2023physics} improved the PINN model, achieving $R^2 \approx 0.9$ and extending the model to bi-dispersed systems by evaluating its performance in simulations containing particles of two different sizes. 
Within the present manuscript, we refer to PINN, MPP and PIEP as Pairwise-Interaction Based (PIB) models. 
Unlike PIB approaches, the FCNN architecture used in the present work does not utilize the assumption of pairwise interaction between particles. In PIB, the interaction between a reference particle and each neighboring particle is computed separately within individual network blocks. The total force on the reference particle is then obtained by summing the contributions from all these blocks. The PIB approach assumes that the forces between two interacting particles lie in a plane defined by their centers and a line parallel to the mean flow, with the resulting torque constrained to an axis perpendicular to this plane. The approach is axisymmetric with respect to the flow direction in the predicted forces. In contrast, our model processes all input parameters simultaneously, without decomposing the force computation into separate pairwise interactions.

Recent advances highlight that data-driven models \cite{seyed2021, cheng2023physics} currently offer the highest precision in predicting hydrodynamic forces, incorporating particle positions and sizes directly into the Neural Network’s (NN) input vector. However, this method may be limited when applied to systems with complex particle geometries and varying sizes, where the input vector must account for both particle positions and geometric characteristics. As particle geometry becomes more intricate, the number of parameters needed to describe these shapes increases significantly. In this work, we propose an alternative approach to construct the neural network input vector. Instead of directly embedding particle positions, we introduce an additional grid to subdivide the domain and compute local particle volume fractions within each grid cell. The information from neighboring cells in this grid is then incorporated into the input vector, enabling the model to predict the hydrodynamic forces on individual particles. \review{As the parameter range for the model, we select Reynolds numbers between 2 and 40 and particle volume fractions in the range $0.1 < \phi < 0.3$, as these values represent flow regimes relevant to sediment transport in moderately rough turbulent channel flows \cite{kidanemariam2017formation,vowinckel2014fluid,scherer2022role,jain2017spanwise,mazzuoli2019direct}.} To develop the data-driven model, we selected the FCNN architecture. \review{Although FCNN may not yield optimal performance, it facilitates straightforward implementation and enables effective model comparison.} This approach allows us to isolate and assess variations in model performance based solely on differences in the construction of the input vector. Although \citet{balachandar2020} and \citet{seyed2021} demonstrated that \review{using particle positions as a direct input for FCNN} can lead to unsatisfactory accuracy with signs of overfitting,  our attempt to use FCNN achieves significantly improved \( R^2 \) scores by leveraging a larger training data set and employing additional optimization techniques.

This article is structured as follows. We first present results from our pr-DNS, with a detailed description of the numerical methodology presented in Section \ref{sec:num_method}. In Section \ref{sec:num_setup}, we detail the computational setup of our simulations. Subsequently, we validate our simulatoins approach in Section \ref{sec:validation} by comparing the averaged and statistical data on hydrodynamic forces with existing averaged models and other numerical results.  We then develop a data-driven model that uses the FCNN architecture as a reference, which incorporates the relative positions of neighboring particles directly, in line with the approach of \citet{he2019supervised}. This model, is referred to as the \PPBlong (PPB) model, and we provide the details of the model and its evaluation in Section \ref{sec:data_driven_model}. Subsequently, we construct the \LVFBlong (\LVFB) model, which also employs the FCNN architecture. In contrast to the \PPB model, the \LVFB model uses volume fraction values computed on an additional NN grid. We assess both the \LVFB and \PPB models in Section \ref{sec:results} by comparing them with other models discussed in the literature. Finally, in Section \ref{sec:conclusion}, we provide the concluding remarks for our study.

\section {Data generation}
\label{sec:data_generation}
\subsection{Numerical method}
\label{sec:num_method}
To generate training and validation data sets for a data-driven model, we perform pr-DNS by solving the unsteady  Navier-Stokes equations for incompressible fluids:
  \begin{equation}
    \label{eq:NS}
    \frac{\partial {\bm{u}}}{\partial t} + \nabla \cdot (\bm{u} \bm{u}) = \frac{1}{\rho_f}\nabla \bm{\tau}_f + \bm{f}_b + \bm{f}_{IBM}
  \end{equation}
  together with the continuity equation 
  \begin{equation}
    \label{eq:continuity}
    \nabla \cdot {\bm{u}} = 0.
  \end{equation}
Here, $\bm{u}$ is the velocity of the fluid, $t$ is the time, $\rho_f$ is the fluid density, and $\bm{\tau}_f=-p\bm{I}+\rho_f\nu\left[\nabla\bm{u} + \left(\nabla \bm{u}\right)^T\right]$ is the fluid stress tensor, $p$ is the fluid pressure with the hydrostatic component subtracted out, $\nu$ is the kinematic viscosity of the fluid, and $\bm{I}$ is the identity \review{tensor}. The term $\bm{f}_b$ represents a pressure gradient throughout the simulation domain to drive the flow. The term $\bm{f}_{IBM}$ originates from the Immersed Boundary Method (IBM) to enforce a no-slip boundary condition on the particles' surface and corresponds to the influence of stationary submerged particles, forcing the velocity to satisfy the no-slip boundary condition on the surface of each particle.
The IBM method was introduced by \citet{uhlmann2005immersed} and further enhanced by \citet{kempe2012improved}. The version of the implementation that we use in the present work is described and validated by \citet{biegert2017collision} and successfully employed  in recent studies \cite{kleischmann2024pairwise,maches2024settling,zhu2024effects}. Equation \ref{eq:NS} is solved numerically in its non-dimensional formulation. By introducing the reference values for a characteristic length scale and velocity, the main variables in equation \ref{eq:NS} are expressed as follows:
  
 \begin{subequations}
    \label{eq:Non_Dim_ref}
    \begin{align}
    \bm{x} = D_p \cdot \Tilde{\bm{x}}, \\
    \bm{u} = u_\infty \cdot \Tilde{\bm{u}}, \\
    t = \frac{D_p}{u_\infty} \cdot \Tilde{t}, \\
    p = \rho_f u_\infty^2 \cdot \Tilde{p},
    \end{align}
\end{subequations}
where $\bm{x} = (x,y,z)^T$ is position vector, the tilde operator ${( \Tilde{\cdot})}$  corresponds to non-dimensional quantities, and $u_\infty$ corresponds to the so-called superficial velocity. The superficial velocity represents the magnitude of the average velocity for the volume of the entire simulation domain \cite{vowinckel2017momentum}.
Inserting these quantities into equations \ref{eq:NS} and \ref{eq:continuity}, and introducing $Re = u_{\infty} D_p/{\nu}$ as the bulk Reynolds number, we reformulate the equation system (Eq. \ref{eq:NS} - \ref{eq:continuity}) in its non-dimensional form:
  
  \begin{equation}
    \label{eq:NS_nondim}
    \frac{\partial {\bm{\Tilde{u}}}}{\partial \Tilde{t}} + \nabla \bm{\Tilde{u}} \bm{\Tilde{u}} = - \nabla {\Tilde{p}} + \frac{1}{Re} \nabla^2 {\bm{\Tilde{u}}} + \bm{\Tilde{f}}_b + \bm{\Tilde{f}}_{IBM},
  \end{equation}
  
  \begin{equation}
    \label{eq:continuity_nondim}
    \nabla {\bm{\Tilde{u}}} = 0.
  \end{equation}

\begin{figure*}[t!]
    \centering
    \includegraphics[width=8cm]{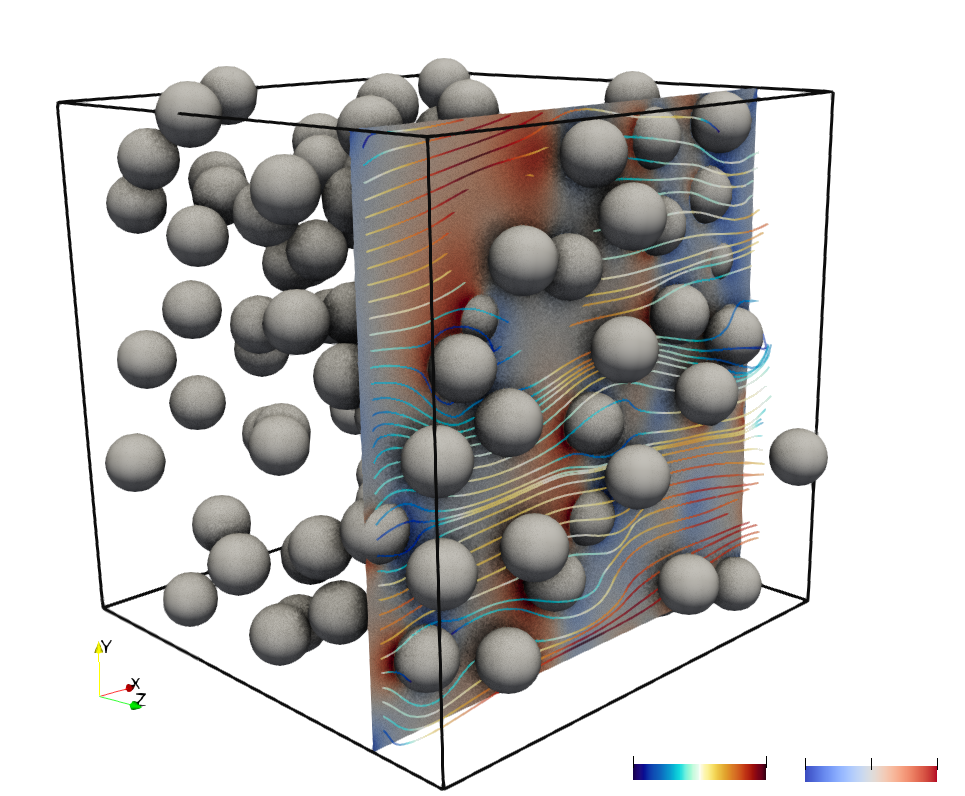}
    \includegraphics[width=8cm]{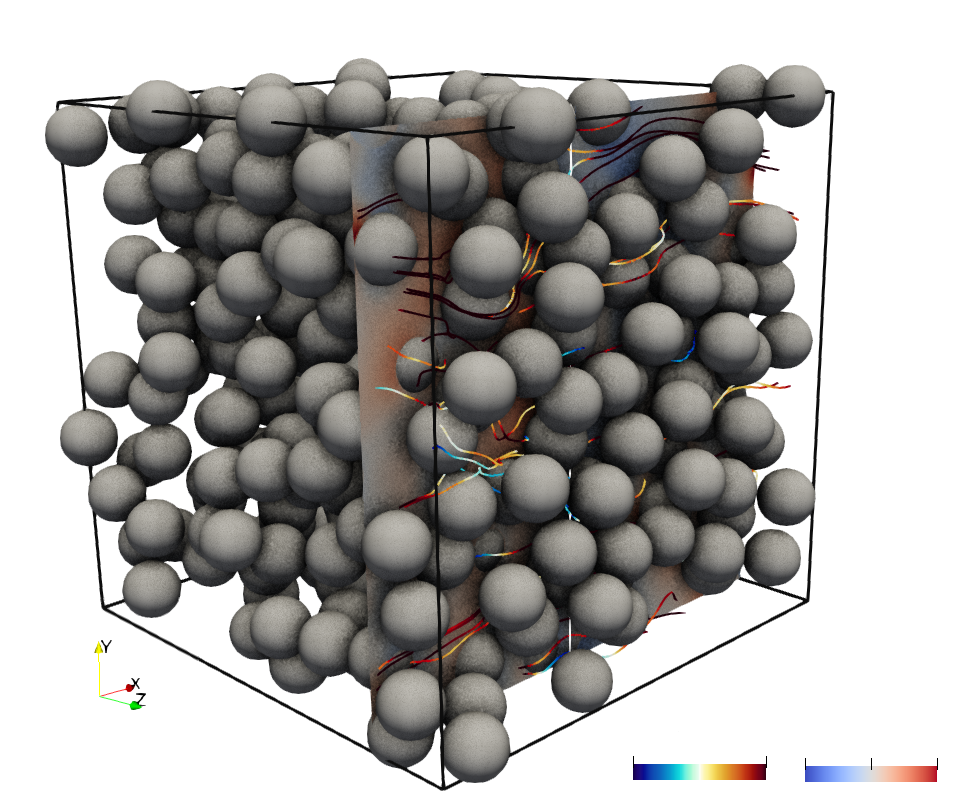}
    \put(-230,160){$b)$}
    \put(-71,20){$||\mathbf{\Tilde{u}}||$}
    \put(-80.5,11){\footnotesize{$0$}}
    \put(-54,11){\footnotesize{$2.5$}}
    \put(-24,20){$\Tilde{p}$}
    \put(-12,11){\footnotesize{$15$}}
    \put(-42,11){\footnotesize{-$15$}}
    \put(-460,160){$a)$}
    \put(-301,20){$||\mathbf{\Tilde{u}}||$}
    \put(-310.5,11){\footnotesize{$0$}}
    \put(-284,11){\footnotesize{$2.5$}}
    \put(-253,20){$\Tilde{p}$}
    \put(-239,11){\footnotesize{$6$}}
    \put(-272,11){\footnotesize{-$6$}}
    \caption{Examples of snapshots of the simulation domain for particle volume fractions $\phi = 0.1$ (a) and $\phi = 0.3$ (b). The slice plane illustrates the distribution of non-dimensional pressure, while the streamlines are colored based on non-dimensional velocity magnitude.}
    \label{fig:domain_illustration}
\end{figure*}
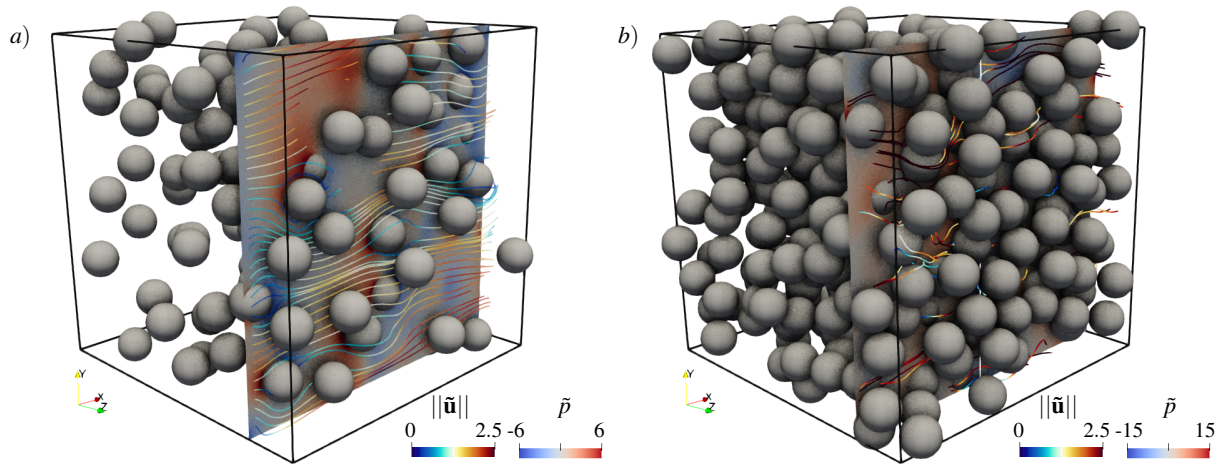

The simulation domain is divided into a Cartesian grid of $N_x \times N_y \times N_z$ uniformly spaced cells, and the variables are arranged using the marker-and-cell method \cite{peric2002cfdbook}. The surface of the particles is modeled by discrete Lagrangian markers, where the no-slip condition is enforced through the IBM \cite{biegert2017collision}. The spatial accuracy of the numerical approach is second-order, as all derivatives are computed using second-order central finite differences. Time integration is performed with a third-order Runge–Kutta method. For the remainder of this article, we continue writing all quantities in a non-dimensional way and drop the tilde for convenience.

\subsection{Numerical setup}
\label{sec:num_setup}

\begin{table}[b]
    \centering
    \begin{tabular}{c|ccccc|ccccc}
     & \multicolumn{5}{c}{Train data set} & \multicolumn{5}{c}{Test data set} \\
     \hline
    \diagbox[width=1cm, height=0.6cm]{ $\phi$ }{$Re$} & 2 & 5 & 10 & 20 & 40 & 2 & 5 & 10 & 20 & 40 \\
    \hline
    0.1 & 250 & 250 & 250 & 250 & 250 & 50 & 50 & 50 & 50 & 50 \\
    0.2 & 250 & 250 & 250 & 250 & 250 & 50 & 50 & 50 & 50 & 50 \\
    0.3 & 160 & 160 & 160 & 160 & 160 & 40 & 40 & 40 & 40 & 40 \\
    \end{tabular}
    \caption{Number of simulations performed for each combination of $Re$ and $\phi$}
    \label{tab:total_simulations}
\end{table}
We perform a series of pr-DNS simulations for a range of combinations of Reynolds numbers $Re \in \{2, 5, 10, 20, 40\}$ and particle volume fractions $\phi \in \{0.1,0.2,0.3\}$. The particle volume fraction $\phi$ is defined as the ratio between the total volume of the number of particles and the volume of the computational domain of the simulations. In order to achieve a certain value of $\phi$, we generate arrangements of monodispersed particles that are fixed in position in space and time and of equal diameter $D_p$ within a cubic computational domain of size $L_x \times L_y \times L_z = 9D_p \times 9D_p \times 9D_p$. Periodic boundary conditions were employed in all directions of the computational domain. \review{The illustration of the computational domain with particle concentration $\phi \in \{0.1,0.3\}$ is presented in Figure \ref{fig:domain_illustration}}. Although a recent study\cite{ELMESTIKAWY2025105143} suggests that such a domain size may be too small to obtain decorrelated velocity fields across the periodic flow direction for \(\phi = 0.05\) and \(Re = 20\), we find that simulations with higher volume fractions (\(\phi > 0.2\)) fulfill this criterion for the same domain size. Hence, our setup is sufficiently large to obtain reliable data. The velocity is initialized as $\mathbf{u}_{init} = \left(1 \quad 0 \quad 0\right)^T$, while the pressure was initialized with zero values $p_{init} = 0$.
The pressure gradient driving the flow $\bm{f}_b$ is applied in the $x$ direction, which we refer to as the streamwise component and calculated in such a way as to enforce the superficial non-dimensional velocity $u_{\infty} = 1$. The resolution of the numerical grid for $\phi \in \{0.1,0.2\}$ is $N_x = N_y = N_z = 250$, while for $\phi = 0.3$ the resolution was refined to $N_x = N_y = N_z = 450$ as we obtain higher flow velocities in the pore space with increasing confinement. This yields a respective resolution of $27.8$ and $50$ grid cells across the particle diameter.

Table \ref{tab:total_simulations} provides a summary of all simulations conducted for the present study. It is seen that for each combination of $Re$ and $\phi$ at least 200 simulations were performed. Each simulation includes between 139 and 417 particles depending on $\phi$. This renders the minimum number of samples for specific $Re$ and $\phi$ to be 41700, which is about one order of magnitude higher than was reported in previous studies \cite{moore2019,he2019supervised,seyed2020,balachandar2020,cheng2023physics}. Full disclosure of this dataset is provided in the supplemental material of this study. The training data set in Table \ref{tab:total_simulations} represents simulations that we used to train the model. Twenty percent of the training data set was designated as the validation set, serving to assess model convergence throughout the training process.

For simulations conducted with volume fractions $\phi \in \{0.1, 0.2\}$, 
particles were placed using a \review{Random Sequential Addition} algorithm that randomly positions them within the domain \cite{widom1966random}. If a particle intersects with any previously placed particle, it is repositioned randomly until no intersections occur. This process repeats until all particles are placed without overlap. The generation of randomly arranged particles with $\phi = 0.3$ using the same distribution algorithm used for $\phi \in \{0.1, 0.2\}$ proved to be computationally inefficient. Consequently, we employed dry discrete element method (DEM) simulations to generate a quasirandom particle distribution more efficiently. \review{The particle dynamics in the DEM simulation are the same as those reported in the work of \citet{zhao2021flocculation}.}
Initially, we distributed particles \review{randomly} throughout the simulation domain. We then applied triple periodic boundary conditions and a gravitational vector in the $y$-direction. Few particles were fixed in place, while others were allowed to move under the influence of gravity. During this dry simulation phase, we sampled the particle positions, ensuring that the sampling frequency was low enough to allow significant particle movement relative to their diameter.

\begin{figure*}[t!]
    \centering
    \includegraphics{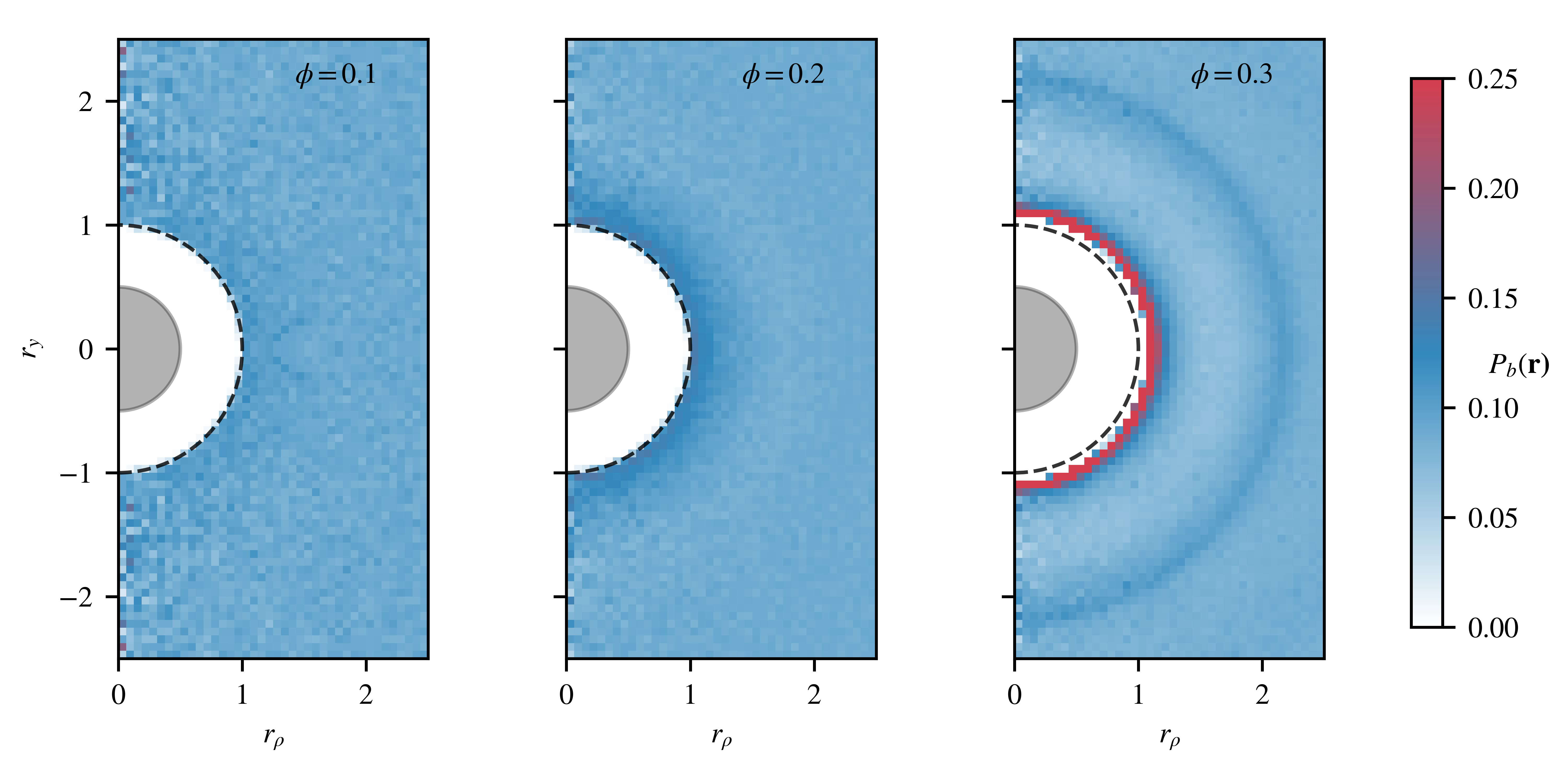}
    \caption{Pairwise particle density distribution $P_b\left(\bm{r}\right)$ for $\phi \in \{0.1,0.2,0.3\}$. The data for the relative positions of the particles pairs were collected in the ranges $0< r_\rho < 2.5$ and $-2.5<r_y<2.5$. The reference particle is shown as a filled gray semicircle, while the outer semicircle indicates the closest possible position of the centers of the neighboring particles. \review{For volume fractions $\phi \in \{0.1, 0.2\}$, particle distributions were generated using the RSA approach, whereas for $\phi = 0.3$, particles were distributed using the DEM simulation of \citet{zhao2021flocculation}}}
    \label{fig:neigh_distribution}
\end{figure*}

To illustrate the distribution of particles for different values of \(\phi\), Figure \ref{fig:neigh_distribution} presents the pairwise particle density distribution function \(P_b(\bm{r})\), which is defined as a function of relative distances between pairs of particles $r$. The function \( P_b(\bm{r}) \) is represented on a two-dimensional uniform rectangular grid, where the subscript \( b \) refers to a specific cell in this grid. We define $\bm{r}$ within two non-dimensional spatial components $y$ and $\rho = \sqrt{x^2+z^2}$. The choice of these components is motivated by the fact that for $\phi=0.3$, the particle distributions were generated by DEM simulations with the gravity vector aligned along the $y$ axis, which can lead to axisymmetric particle distributions around the $y$ axis. We define the relative distances between the particles $i$ and $j$ as $\bm{r}^{i,j}= (r_\rho^{i,j}, r_y^{i,j})^T = (\rho_p^i - \rho_p^j, \ y_p^i - y_p^j)^T $, where $\rho_p$ and $y_p$ are $\rho$ and $y$-components of the particles $i$ and $j$, respectively.  The pairwise particle density distribution function \(P_b(\bm{r})\) is calculated within $0< r_\rho < 2.5$ and $-2.5<r_y<2.5$ and defined as 
\begin{equation}
\label{eq:pairwise_distribution}
    P_b(\bm{r}) = \frac{1}{\Delta b^2 N_\phi n(n-1)}\sum_{k}^{N_\phi}\sum_{i}^n\sum_{j \neq i}^n K\left(\frac{||\bm{r}^b - \bm{r}^{i,j}||_\infty}{\Delta b}\right)
\end{equation}
such as the total probability of finding a neighbor in the area $S = \{ (r_\rho, r_y) \mid 0 < r_\rho < 2.5, \ -2.5 < r_y < 2.5 \}$ is $\int P_b(\bm{r}) dS = 1$. Here, $\Delta b = 1/16$ is the non-dimensional size of a grid cell used for this analysis, $N_\phi$ is the number of train and test simulations for a specific $\phi$, $n$ is the number of particles per domain, and $\bm{r}^b$ is the center of the 2D grid cell $b$. The kernel $K(u)$ is defined as 
\begin{equation}
\label{eq:kernel}
    K(u) =
    \begin{cases} 
    1, & \text{if } |u| < 1/2, \\
    0, & \text{otherwise}.
    \end{cases}
\end{equation}
A perfectly uniform distribution of particles would correspond to a constant probability density function \( P_b(\bm{r}) = 1/S =0.08 \), indicating a uniform likelihood of finding a neighboring particle at any location \( \bm{r} \). However, particles cannot overlap, meaning the center of any neighboring particle cannot be located closer than one particle diameter from the center of another. This exclusion leads to a modification of \( P_b(\bm{r}) \), such as in the immediate vicinity of a particle, the probability is zero within the exclusion zone \( ||\bm{r}||_2 < D_p \), and the distribution becomes non-uniform. At higher volume fractions, this constraint causes particles to be more tightly packed beyond the exclusion zone, which in turn increases the likelihood of finding a neighbor just outside one diameter away. As a result, \( P_b(\bm{r}) \) exhibits peaks at distances slightly beyond the particle diameter, and this effect becomes more pronounced as the volume fraction increases.
The gray filled semicircle with the center at $(r_\rho, r_y)^T = (0,0)^T$ in Figure \ref{fig:neigh_distribution} represents the reference particle, while the dashed black semicircle represents the closest possible location of the center of neighboring particles. 
As seen in Figure \ref{fig:neigh_distribution}, an increase in volume fraction $\phi$ leads to stronger spatial correlations and more structured particle arrangements around the reference particle. At $\phi=0.1$, the density distribution is more uniform, with no visible increase of $P_b(\bm{r})$ within the vicinity of the reference particle. However, at $\phi=0.2$ and $\phi=0.3$, the values of $P_b(\bm{r})$ increases significantly near the reference particle, suggesting a closer clustering. For $\phi=0.3$, the highest density is observed in the immediate vicinity of the reference particle, with a second peak emerging at $||\bm{r}||_2 \approx 2$. The second peak is attributed to the enhanced particle packing and layering effects caused by the higher particle concentration. It is important to note that the first peak of the density distribution does not occur exactly at $ ||\bm{r}||_2 = 1$, but further away. This shift is a result of the DEM simulations, where particles were modeled with slightly larger radii to avoid particle overlap due to soft-sphere collisions. Despite randomization of the positions of the particles in $\phi=0.3$, no discernible microstructural patterns are observed in either the vertical or horizontal directions.

\begin{figure*}[t!]
    \centering
    \includegraphics{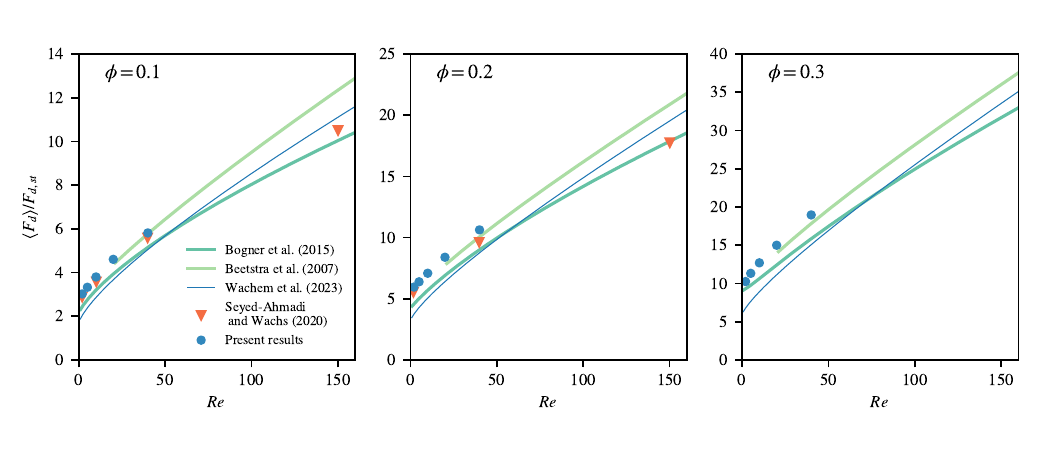}
    \caption{Averaged drag force $\langle F_d \rangle$ normalized over the Stokes drag $F_{d,st}$ for different particle volume fractions compared with models of Bogner et al. and Beestra et al. \cite{bogner2015,beetstra2007drag}}
    \label{fig:validation}
\end{figure*}

\subsection{Validation}
\label{sec:validation}
We validated our numerical simulations by comparing the mean values of the hydrodynamic forces acting on the particles and their respective variances to studies that performed similar numerical simulations. Figure \ref{fig:validation} presents a comparison of the streamwise drag force averaged over all simulations $\langle F_x \rangle$ as a function of the Reynolds number for $\phi \in \{0.1,0.2,0.3\}$. Here, the angle bracket operator $\langle \cdot \rangle$ denotes the averaging over all particles within all simulation cases with specified values of $Re$ and $\phi$. The average drag force is normalized by the Stokes drag force for a single particle in uniform flow $F_{d,St} = 3\pi\mu D_p u_{\infty}$, where $\mu$ is the dynamic viscosity of the ambient fluid.

We also refer to the $y$ and $z$ components of the hydrodynamic force (i.e. $F_y$ and $F_z$) as the lateral (or lift) component $F_l$.
Consequently, we define the average lift force as $\langle F_l \rangle = \frac{1}{2}(\langle F_y \rangle + \langle F_z \rangle)$. In Sections \ref{sec:ppb_performance} and \ref{sec:lvfb_performance}, we report the \( R^2 \) score for the lift force, which we calculate as the average of the \( R^2 \) scores for \( F_y \) and \( F_z \). We do not use the magnitude of the lateral force (i.e. $(F_y^2 + F_z^2)^{1/2}$) to be consistent with previous studies \cite{seyed2020,seyed2021,cheng2023physics}, where the lateral component of the force was projected onto the $y$ or $z$ axis. Similarly, we define the lateral torque, \( T_l \), as the average of the \( y \) and \( z \) components of the hydrodynamic torque, and compute its \( R^2 \) score as the average of the \( R^2 \) scores for \( T_y \) and \( T_z \), following the same approach used for \( F_l \).

The simulation results are evaluated against empirical correlations, which are derived from multiple numerical studies. The numerical methods used in these studies include the Lattice Boltzmann Method (LBM), as used by \citet{beetstra2007drag} and \citet{bogner2015}, and the IBM, which is applied in the studies of \citet{yali2015new}, \citet{seyed2020}, and \citet{van2023microstructure}.  As can be seen in Figure \ref{fig:validation}, the results of our simulations strongly correlate with the model derived by \citet{beetstra2007drag} and are very similar to the results of \citet{seyed2020} for $\phi \in \{0.1,0.2\}$. The exact values of the normalized average drag force $\langle F_x \rangle / F_{d,St}$ are provided in Table \ref{tab:force_statistics}.

\begin{table}[]
    \centering
    \begin{tabular}{rrcccccc}
    $\phi$ & $\; Re \;$ & $\frac{\;\langle F_x \rangle}{F_{{d,st}}} \;$ & $\frac{\sigma_{F_x}}{\langle F_x \rangle}$ & \review{$\left( \frac{\sigma_{F_x}}{\langle F_x \rangle} \right)_{sw}$} & $\frac{\langle F_l \rangle}{F_{{d,st}}}$ & $\frac{\sigma_{F_l}}{\langle F_x \rangle}$ & \review{$\left( \frac{\sigma_{F_l}}{\langle F_x \rangle}\right)_{sw}$} \\
    \toprule
     & 2 & 3.02 & 22.4 \% & 21.3 \% & 1.3e-07 & 14.3 \% & 15.1 \% \\
     & 5 & 3.33 & 23.5 \% & - & -3.6e-05 & 14.6 \% & - \\
    0.1 & 10 & 3.80 & 25.6 \% & 24.8 \% & 7.1e-05 & 14.5 \% & 14.8 \% \\
     & 20 & 4.61 & 26.8 \% & - & 5.1e-04 & 14.7 \% & - \\
     & 40 & 5.82 & 29.0 \% & - & 1.7e-04 & 14.9 \% & - \\
     \hline
     & 2 & 5.96 & 19.8 \% & 17.6 \% & 2.4e-12 & 16.0 \% & 15.2 \% \\
     & 5 & 6.39 & 20.8 \% & - & 5.8e-08 & 16.1 \% & - \\
    0.2 & 10 & 7.09 & 22.4 \% & - & 2.2e-06 & 16.5 \% & - \\
     & 20 & 8.40 & 24.2 \% & - & -2.8e-05 & 16.8 \% & - \\
     & 40 & 10.62 & 25.7 \% & 23.8 \% & -3.7e-06 & 17.3 \% & 16.3 \% \\
     \hline
     & 2 & 10.25 & 26.3 \% & - & 1.1e-08 & 22.1 \% & - \\
     & 5 & 11.31 & 27.1 \% & - & 1.5e-07 & 21.7 \% & - \\
    0.3 & 10 & 12.70 & 27.2 \% & - & -3.9e-07 & 21.6 \% & - \\
     & 20 & 15.03 & 28.1 \% & - & 1.6e-06 & 20.0 \% & - \\
     & 40 & 18.97 & 28.8 \% & - & -1.7e-04 & 18.9 \% & - \\
    \end{tabular}
\caption{Force statistics over all performed simulations sorted by the Reynolds number and particle volume fraction. \review{Quantities with the subscript $sw$ refer to data reported by \citet{seyed2020}.}}
\label{tab:force_statistics}
\end{table}

We also examine the distributions of drag and lift forces. Table \ref{tab:force_statistics} provides the corresponding variances of the drag and lift forces for all simulations with different $\phi$ and $Re$. Within this analysis, we found a very close agreement of our data with the simulation results of \citet{seyed2020}, \review{which are also included in Table \ref{tab:force_statistics} under the subscript $sw$}. Similarly to \citet{seyed2020}, the standard deviation of the drag force divided by the mean drag force $\sigma_{F_x} / \langle F_x \rangle$ lies in a range between 20 and 30 percent. The deviation of the lift forces divided by the mean drag force is in the range between $14 \% < \sigma_{F_l} / \langle F_x \rangle <  22 \%$, which is very close to the range reported by \citet{seyed2020} ($14 \% < \sigma_{F_l} / \langle F_x \rangle <  20 \%$). Lastly, we observed negligibly small values of $\langle F_l \rangle / F_{d,St}$, indicating that there is no preferential direction for the lateral component of the hydrodynamic force. Hence, we conclude that particles are not arranged in a way that biases the flow towards any specific tangential direction.

\section{Data driven model}
\label{sec:data_driven_model}
To construct a data-driven model we employ FCNN architecture to predict hydrodynamic forces and torques on particles. While previous studies have shown that FCNNs offer limited applicapility for this task \cite{he2019supervised,balachandar2020,seyed2021}, in the present work, we demonstrate that FCNN performance can be significantly enhanced beyond what was previously reported.
As shown in Section \ref{sec:ppb_performance}, increasing the size of the data set in conjunction with regularization techniques such as dropout, batch normalization, and training methods such as Cyclic Learning Rate (CLR) \cite{smith2017cyclical}, leads to a substantial improvement in the accuracy of the model compared to previously reported FCNN models. Nevertheless, the performance of the model remains inferior to that of the previously reported PIB models. Section \ref{sec:lvfb_performance} further demonstrates that the use of the same data set while modifying the input layer construction methodology and incorporating the data augmentation technique increases the accuracy of the FCNN model and makes it comparable to the PINN models reported by \citet{seyed2021}. The foundational definitions and common training procedures for both the \PPB and \LVFB models are outlined in Section \ref{sec:general_model_formulation}.

\subsection{Model structures and tuning approach}
\label{sec:general_model_formulation}

\begin{figure}[!b]
    \centering
    \includegraphics{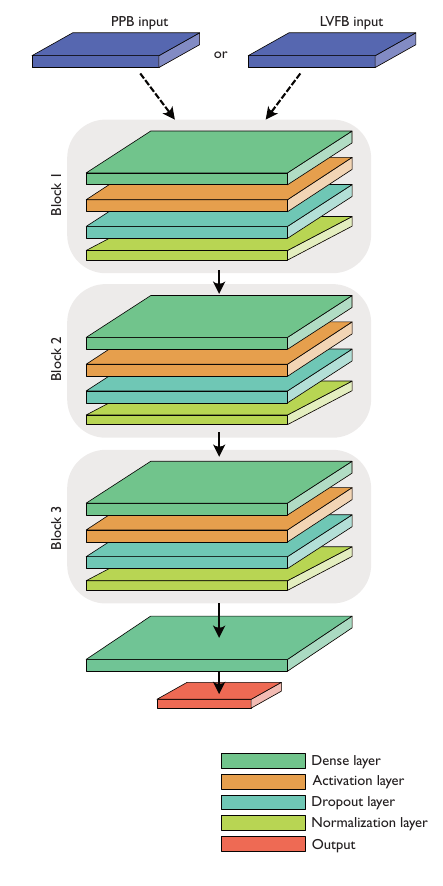}
    \caption{Schematics of the fully-connected NN architecture used in the present work for \LVFB and \PPB methods}
    \label{fig:schematics_FCNN_CN}
\end{figure}

Figure \ref{fig:schematics_FCNN_CN} shows the main structure of the NN architecture used in the present study, which consists of several components. First, the input layer is constructed differently for the \PPB and \LVFB methods. We provide further details on how the input vectors are constructed in Sections \ref{sec:ppb_method} and \ref{sec:lvfb_method} below. Although the input vectors are constructed differently for both approaches, we process input data using a scaling technique such as the Maximum Absolute Scaler 
to ensure that all features are on the same scale. The scaled inputs are then fed into the NN consisting of three fully connected blocks. Each block contains a fully connected layer, an activation function to introduce nonlinearity, a dropout layer for regularization, and a batch normalization layer to stabilize training. After these blocks, a final fully connected output layer produces the predicted forces and torques acting on the reference particle and maps it onto the output. The sizes of fully connected layers between the blocks might differ within the range indicated as "Neurons per hidden layer" in Table \ref{tab:effective_configurations}.

\begin{table}[t!]
    \centering
    \begin{tabular}{l|l}
        \hline
        \textbf{Parameter} & \textbf{Tested Range} \\
        \hline
        Neurons per hidden layer & 64 – 1024 \\ 
        Activation function & [ReLU, Swish, GELU, PReLU]\\
        Dropout Rate & 0 – 0.3 \\
        Scaler & [MaxAbsScaler, RobustScaler, \\  
        & StandardScaler, PowerTransformer] \\
        Batch Size & 128 – 8192 \\
        Optimizer & [NAdam, Adam, RectifiedAdam,\\
        & Yogi] \\
        Loss Function & [MSE, MAE] \\
        Learning Rate & 0.0001 – 0.05 \\
        CLR Mode & [Exponential Decay, Triangular,\\ 
        & Triangular2]\\
        Peak Learning Rate for CLR & 0.05 – 1 \\
        \hline
    \end{tabular}
    \caption{Hyperparameters and their corresponding ranges tested for \PPB and \LVFB models.}
    \label{tab:effective_configurations}
\end{table}

Hyperparameters are critical settings that define the architecture and training behavior of the model but are not learned during training. After constructing the main architecture of the model, we performed hyperparameter tuning by exploring different combinations of hyperparameters to identify effective configurations. The comprehensive list of the hyperparameters adjusted in this study is provided in Table \ref{tab:effective_configurations}, where the corresponding tuning ranges are also presented. For a more complete description of various optimizers and CLR modes mentioned in Table \ref{tab:effective_configurations}, readers are referred to the official TensorFlow documentation (https://www.tensorflow.org). Similarly, detailed information on the scalers listed in Table \ref{tab:effective_configurations} can be found in the scikit-learn library documentation (https://scikit-learn.org/). 

To explore the hyperparameter space faster, the models were trained in parallel using different hyperparameter configurations. This approach significantly reduced the time required to identify an optimal model by allowing us to evaluate many candidate models at once, rather than sequentially testing each configuration. For the actual optimization process, we used Bayesian optimization\cite{BayesianOptimization}, which intelligently selects hyperparameter values based on past results rather than testing them randomly. This method builds a probabilistic model to predict which hyperparameter sets are likely to perform well, thus focusing computational resources on promising configurations. In addition, we incorporated a hyperband pruner\cite{hyperbandPruner} to terminate models that showed poor performance in the early stages of training. This prevented unnecessary computation on models that can be expected to perform poorly, allowing us to allocate more resources to more promising candidates. 
Hyperparameter tuning revealed that using the Maximum Absolute Scaler in combination with the "NAdam"\cite{NAdam} optimizer consistently yielded the best-performing models. The optimal number of neurons per hidden layer was found to be 512 in most of these best performing models. Furthermore, the use of CLR was found to enhance the convergence rate of the models.

We identified the best-performing model configurations by evaluating their coefficients of determination (or $R^2$ scores). We calculate $R^2$ scores as following: 
\begin{equation}
\label{eq:r2_score}
    R^2 = 1 - \frac{\sum_i^\mathcal{T} (Q_i^{Pr} - Q_i^{DNS})}{\sum_i^\mathcal{T} (Q_i^{Pr} - \langle Q_i^{DNS} \rangle)}
\end{equation}
where, $Q$ corresponds to a specific component of force or torque, superscript $Pr$ corresponds to the force or torque predicted by the model, and the superscript $DNS$ corresponds to the force or torques calculated as the results of pr-DNS. The summation set \( \mathcal{T} \) represents particles from a specific set of simulations, and the index $i$ refers to the individual particle. The 20 models achieving the highest $R^2$ scores with $Q = F_x$ for the entire validation data set \( \mathcal{T} = \mathcal{T}_{test} \) were then selected for further evaluation. Then these models were evaluated based on their \( R^2 \) scores for different subsets of the validation data set, corresponding to various combinations of \( Re \) and \( \phi \). The subsets are denoted as \( \mathcal{T}_{test}^{Re, \phi} \subseteq \mathcal{T}_{test} \), and the evaluation was performed for \( Q \in \{F_x, F_l, T_l\} \). After this evaluation, the optimal hyperparameters for the \PPB and \LVFB models were identified. The results of these selected approaches are presented in Section \ref{sec:results}.

\subsection{The \PPBlong (PPB) model}
\label{sec:ppb_method}
Within the \PPBlong model we follow an approach similar to \citet{he2019supervised} by including the Reynolds number ($Re$), volume fraction ($\phi$), and relative positions of neighboring particles $i$ to a reference particle $j$ defined in Cartesian coordinates (i.e. $\bm{r}^{i,j} = (r_x^{i,j},r_y^{i,j},r_z^{i,j})^T$) in the input vector. However, unlike \citet{he2019supervised}, who selected a fixed number of closest-neighboring particles, we defined a spherical region centered on the particle of interest with a radius $R_{eff}$. Particles whose centers lie within this spherical region are included in the input vector. Given that the number of neighboring particles varies depending on the local packing density, we designed the input vector big enough to accommodate the maximum possible number of neighbors for the highest possible spherical packing fraction. If the number of particles within the defined region is less than the input vector size, we initialize the remaining values to zero, which is typically referred to as zero padding. 
This approach allows us to consider different numbers of particles in the defined area around the particle of interest.
The output vector of the model contains the total hydrodynamic forces and torques, \((F_x, F_y, F_z, T_x, T_y, T_z)^T\). This differs from the models presented in \cite{seyed2020,seyed2021,balachandar2020,cheng2023physics}, where only the fluctuating components of the hydrodynamic forces and torques are predicted.

\subsection{The \LVFBlong (LVFB) model}
\label{sec:lvfb_method}

\begin{figure}[t]
    \centering
    \includegraphics{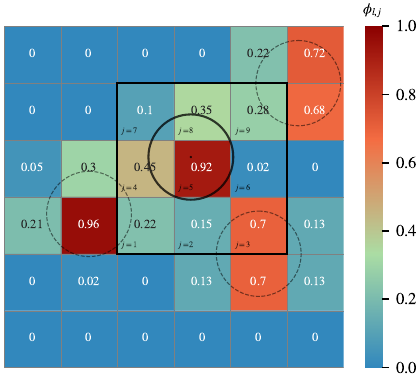}
    \put(-95.5,145){\vector(-1,0){50}}
    \put(-114,145){\vector(1,0){50}}
    \put(-117,148){\scriptsize{$L_{eff}$}}
    \put(-150,120){\vector(0,1){20}}
    \put(-150,133.5){\vector(0,-11){20}}
    \put(-155.5,125){\scriptsize{$L$}}
    
    \caption{2D-illustration of the local volume fraction $\phi_{l}$ within \textit{NN grid}. The coloring of \textit{NN cells} correspond to the values of local volume fraction $\phi_{l}$, the values of which are also shown within each \textit{NN cell}. The solid circle corresponds to the reference particle, for which hydrodynamic forces are predicted and dashed circles correspond to neighboring particles. Black rectangle represents the area of influence\review{, which is determined by considering only one adjacent cell (i.e. $W=1$)}.}
    \label{fig:vof_illustration}
\end{figure}
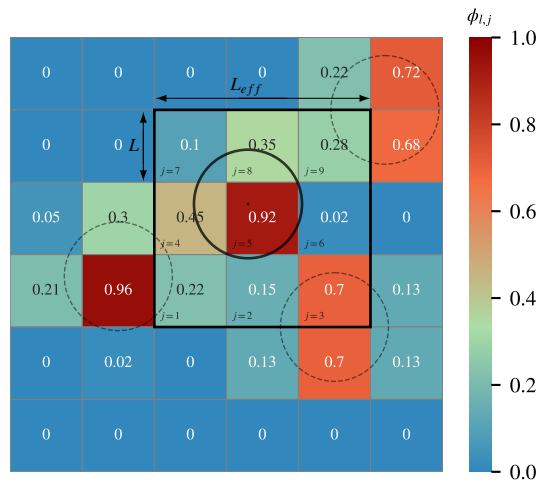

In contrast to the \PPB model, we propose an alternative approach to constructing the input vector. This method computes the local particle volume fraction on an auxiliary grid that is significantly coarser than the pr-DNS simulation grid, but finer than the simulation grid that is used within the point-particle (or mesoscale) simulation approach. We refer to this coarser grid as the \textit{NN grid}. The cell of the \textit{NN grid} is also referred to as \textit{NN cell}. Figure \ref{fig:vof_illustration} provides a 2D schematic representation of this method, where the domain is divided by the \textit{NN grid} (gray rectangular mesh) and within each \textit{NN cell} the local particle volume fraction is calculated. The particle for which hydrodynamic forces and torques are predicted is referred to as the reference particle. The reference particle is shown as a solid black circle in Figure \ref{fig:vof_illustration}, while the surrounding particles are shown as dashed circles. To predict the hydrodynamic forces and torques in the reference particle, the local volume fractions ($\phi_l$) from the nearby \textit{NN cells} are used within the input vector of the model. The volume that encompasses all the \textit{NN cells} incorporated within the input vector is called the volume of interest. In Figure \ref{fig:vof_illustration} the volume of interest is enclosed by the solid square and contains nine \textit{NN cells}. We define the volume of interest as a cuboidal subset of the \textit{NN cells} adjacent to the cell, within which the center of reference particle is located. The number of adjacent layers is determined by the variable $W$, which equals one for the illustration shown in Figure \ref{fig:vof_illustration}. Together with the length of the \textit{NN cell} ($L$), the effective length of the volume of interest is defined as $L_{eff} = L(1+2W)$.The variables $L$ and $L_{eff}$ are also shown in Figure \ref{fig:vof_illustration}.

The input vector for the \LVFB method corresponds to the following: $(Re, \phi, X_c, Y_c, Z_c, \phi_{l,1},.., \phi_{l,j},.., \phi_{l,n})^T$. The variables $X_c$, $Y_c$, and $Z_c$ denote the relative position of the reference particle with respect to the center of the \textit{NN cell} in which it is located. The terms $\phi_{l,1}$,.., $\phi_{l,j}$,.., $\phi_{l,n}$ correspond to volume fractions within each \textit{NN cell} within the volume of interest. For the 2D example, shown in Figure \ref{fig:vof_illustration}, the values of the local volume fractions are equal $\phi_{l,1} = 0.22$, $\phi_{l,2} = 0.15$, $\phi_{l,3} = 0.7$ and so on. The total number of local volume fractions $\phi_{l,n}$ included in the input vector is indicated by the subscript $n$ and is determined by $W$. For example, in a 3D domain, $W=1$ would correspond to 27 adjacent \textit{NN cells}, and $W=2$ corresponds to 125 adjacent \textit{NN cells}.  Note that the local volume fraction of the reference particle itself is also included in the calculation of $\phi_l$.

The modified structure of the newly proposed LVFB-method also allows us to use data augmentation to enhance our training data set beyond the originally simulated cases. This is achieved by slightly translating all particles within the domain. As a result, the recalculated volume fractions on the \textit{NN grid} change together with the relative positions of the particles within the \textit{NN cells}. At the same time, the overall arrangement of the particles in relation to each other remains unchanged. Since local flow conditions are preserved, the forces acting on the particles remain the same as in the original data set. As a result, the model gains a broader representation of the input variations, enhancing its ability to distinguish relevant features with minimal additional computational cost.

A disadvantage of the \LVFB method compared to the \PPB is related to the additional computational cost associated with the \textit{NN grid}. The size of \textit{NN grid} is determined by $(H/L)^3$, where $H$ represents the reference size of the domain. Consequently, the storage requirement for the \textit{NN grid} also scales as $(H/L)^3$. Once the \textit{NN grid} is initialized, the volume fraction values must be computed for each \textit{NN cell}. This can be achieved by iterating over all particles within the domain. For each iteration, the \textit{NN cells} intersecting a particle are identified, and the intersection volumes between the particle and these \textit{NN cells} are calculated and assigned to each \textit{NN cell}  accordingly. In the present work, we calculated the intersection volumes between particles and \textit{NN grid} using the algorithm described by \citet{STROBL2016158}. The number of \textit{NN cells} that a particle can intersect ($k$) depends on the ratio $L/D_p$. When $L/D_p = 1$, a particle can intersect with up to 8 cells ($k=8$), while for $L/D_p = 0.25$, the number of possible intersections increases to 125 ($k=125$). If multiple particles intersect with the same \textit{NN cell}, their respective intersection volumes accumulate within that cell. After processing all particles, the total intersection volumes assigned to each \textit{NN cell} are divided by the cell volume to determine the local volume fraction $\phi$. In general, the computational complexity of determining the local volume fraction is $O(kn)$, which means that it scales linearly with both the number of particles in the simulation domain ($n$) and the number of possible intersections ($k$) for each particle.

In contrast to the \LVFB approach, the \PPB approach does not construct the additional mesh but requires the identification of $N_{p}$ nearest neighbors for each particle. This can be done efficiently using the domain-decomposition method. The method involves dividing the domain into rectangular subdomains. Within the domain decomposition method, each particle only considers other particles from its own subdomain and the adjacent subdomains to find the nearest neighbors. The complexity of this approach depends on the number of subdomains and the average number of particles per subdomain. For a well-balanced decomposition, the cost can be approximated as $O(nN_p)$, where $n$ accounts for the loop over all particles in the domain and $N_p$ arises from querying neighboring particles across adjacent subdomains. This approach is particularly efficient when the particle distribution is relatively uniform and the number of subdomains is chosen appropriately to minimize overlap. 

By comparing the amount of computational calculations required to prepare the input data for both the \PPB and \LVFB models, it can be concluded that the computational scaling of both models is comparable and practically depends on the size of \textit{NN cells} for the \LVFB model and the number of neighboring particles considered for the \PPB model. In terms of memory costs, the \LVFB approach is more demanding compared to \PPB as it requires to store the grid with locally calculated volume fractions $\phi$.

\begin{figure*}[!]
    \centering
    \includegraphics{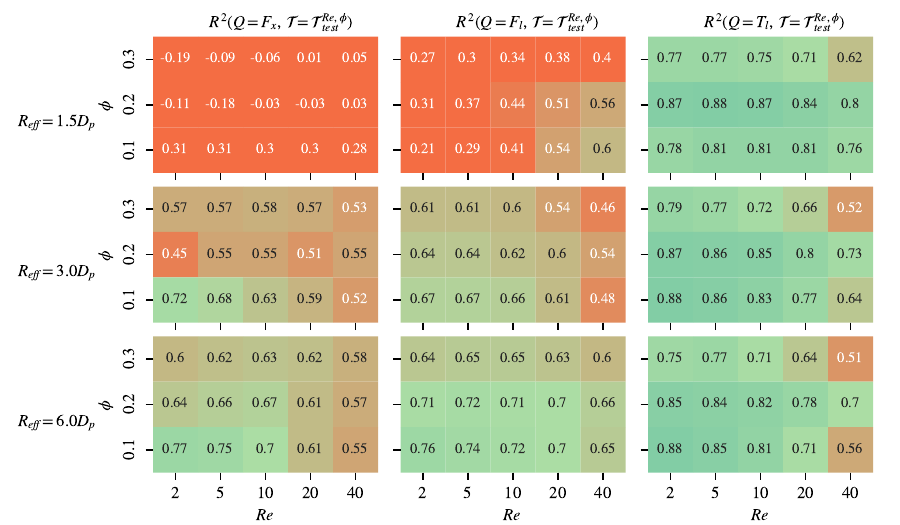}
    \caption{$R^2$ scores for three best \PPB models that correspond to specific $R_{eff}$. The $R^2$ scores are specified for each $Re$ and $\phi$ combination. The color scheme applied to the tables corresponds to the $R^2$ values, with orange indicating lower $R^2$ scores and green representing higher $R^2$ values.}
    \label{fig:NP_best_models_tables}
\end{figure*}

\section{Results}
\label{sec:results}
\subsection{Performance of the \PPB model}
\label{sec:ppb_performance}
We assess the performance of the models using the $R^2$ scores (see Eq. \ref{eq:r2_score}) with $Q \in \{F_x, F_l, T_l\}$ and $\mathcal{T} = \mathcal{T}_{test}^{Re, \phi}$ which is consistent with previous studies \cite{seyed2020,seyed2021,moore2019,cheng2023physics,balachandar2020}. To analyze the sensitivity of the effective radius ($R_{eff}$), defining the range within which neighboring particles contribute to the input vector, we trained models with varying values of $R_{eff} = \in \{1.5D_p, 3.0D_p, 6.0D_p\}$. Naturally, increasing $R_{eff}$ raises computational costs due to the potential inclusion of additional particle data. At the same time, a larger search radius may improve model accuracy by incorporating more information on the positions of neighboring particles. Figure \ref{fig:NP_best_models_tables} presents the $R^2$ scores for the hydrodynamic forces and torque, comparing the pr-DNS results with predictions of models trained using different $R_{eff}$ values. Note that these $R^2$ scores correspond to the test data set, which was not used during training. As expected, the model's predictive accuracy improves as $R_{eff}$ increases, reflecting the incorporation of a larger number of neighboring particles. The highest $R^2$ values were obtained with $R_{eff} = 6.0D_p$. There is also a clear trend in the predictions: lower Reynolds numbers ($Re$) and volume fractions ($\phi$) yield mainly higher $R^2$ values. For $R_{eff} = 6 D_p$, the maximum $R^2$ for the drag force ($R^2 = 0.77$) occurs at $Re = 2$ and $\phi = 0.1$, while the minimum ($R^2 = 0.55$) is observed at $Re = 40$ and $\phi = 0.1$. Although these $R^2$ scores are lower compared to the PINN model by \citet{seyed2021}, where $R^2 = 0.81$ for $\phi = 0.1$, $Re = 2$ and $R^2 = 0.7$ for $\phi = 0.1$, $Re = 40$, they are significantly higher than those reported for the FCNN model by \citet{balachandar2020}. The study by \citet{balachandar2020} reported only negative $R^2$ scores for the drag force and $R^2 < 0.13$ for the lift force. Furthermore, the precision of the present model is comparable to the PIEP model proposed by \citet{balachandar2020}, which achieved $R^2$ scores between 0.65 and 0.75 for drag force with $Re \in \{16,39\}$ and $\phi \in \{0.1,0.4\}$. Therefore, it can be concluded that when trained on large enough data sets and employing the additional optimization techniques such as CLR, dropout, and batch normalization, the FCNN-based model does not suffer from overfitting and is capable of achieving significantly higher accuracy than previously considered.

\begin{figure*}[t!]
    \centering
    \includegraphics{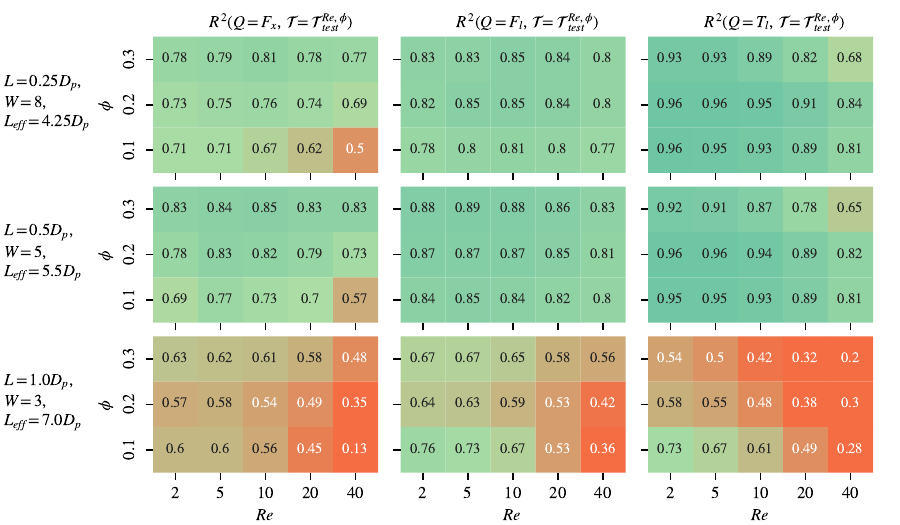}
    \caption{$R^2$ scores for three best \LVFB models that correspond to specific $L$ and $W$. The $R^2$ scores are specified for each $Re$ and $\phi$ combination. The color scheme applied to the tables corresponds to the $R^2$ values, with orange indicating lower $R^2$ scores and green representing higher $R^2$ values.}
    \label{fig:VF_best_models_tables}
\end{figure*}

\subsection{Performance of the \LVFB model}
\label{sec:lvfb_performance}

\begin{table}[b!]
    \centering
    \begin{tabular}{lllcccc}
    $L$ & $W$ & $L_{eff}$ & $R^2(F_x)$ & $R^2(F_l)$ & $R^2(T_l)$ & $R^2 (F_x, \mathcal{T}_{test}^{Re=10, \phi=0.1})$ \\
    \hline
    \hline
    \multirow{2}{*}{$0.25D_p$}& 6 & $3.25D_p$ & 0.96 & 0.75 & 0.93 & 0.58 \\
    & 8 & $4.25D_p$ & 0.97 & 0.83 & 0.94 & 0.67 \\
    \hline
    \multirow{4}{*}{$0.5D_p$} & 2 & $2.5D_p$ & 0.9 & 0.54 & 0.88 & 0.38 \\
     & 3 & $3.5D_p$ & 0.96 & 0.75 & 0.86 & 0.58 \\
     & 4 & $4.5D_p$ & 0.97 & 0.78 & 0.89 & 0.66 \\
     & 5 & $5.5D_p$ & 0.98 & 0.88 & 0.93 & 0.73 \\
    \hline
    \multirow{2}{*}{$D_p$} & 2 & $5D_p$ & 0.96 & 0.68 & 0.6 & 0.63 \\
     & 3 & $7D_p$ & 0.96 & 0.67 & 0.55 & 0.56 \\
    \end{tabular}
    \caption{$R^2$ scores of trained \LVFB models. The columns \( R^2(F_x) \), \( R^2(F_l) \), and \( R^2(T_l) \) were computed using the set of particles from the entire test data set $\mathcal{T}_{test}$.
    }
    \label{tab:various_L_W_parameters}
\end{table}

To optimize the precision of this \LVFB approach, we trained its models for different parameter values of $L$ and $W$ to find their optimal values. \review{On the one hand, smaller $L$, provides more detailed information on the particle distribution, which potentially results in a higher accuracy of the model;} however, for smaller $L$, more memory is needed to store the \textit{NN grid}, and additional computational resources are required to compute the local volume fractions ($\phi_l$). \review{Moreover, decreasing $L$ leads to an increase of the model's input layer size, which also affects the prediction time. For example, with similar $L_{\text{eff}} \approx 5D_p$, reducing $L$ from $D_p$ to $0.1D_p$ increases the number of input variables from $\sim10^2$ to $\sim10^5$. To balance resolution with computational and memory demands, we set $L = 0.25D_p$ as the minimum in this study. The parameter $W$ was selected to ensure that the effective volume around a reference particle captures sufficient information about its neighbors while keeping the model size manageable. Therefore, selecting the appropriate values for both $L$ and $W$ involves balancing the accuracy of the model with the usage of memory and the computational cost for predictions.}

The results of the overall performance of the model are presented in Table \ref{tab:various_L_W_parameters}. We tested three values for the length of the subdomain $L \in \{0.25 D_p, 0.5 D_p, D_p\}$ and also various values of $W$. Table \ref{tab:various_L_W_parameters} shows the $R^2$ scores for the drag and lift forces ($F_x$ and $F_l$, respectively) and for the lateral component of the torque ($T_l$) for the entire test data set. Table \ref{tab:various_L_W_parameters} also provides the $R^2$ score for the drag force of the test data set, which corresponds to $Re = 10$ and $\phi = 0.1$. Although the $R^2$ scores that correspond to the entire test data set provide an overall estimate of the performance of the model, the more accurate representation of the actual precision of the model is determined by the $R^2$ scores for individual combinations of $Re$ and $\phi$, which we provide later in the manuscript. It can be seen from Table \ref{tab:various_L_W_parameters} that the model with $L=D_p$ has the lowest accuracy in terms of $F_l$ and $T_l$ compared to the models with $L= 0.25 D_p$ and $L=0.5 D_p$. Although the length of the domain of interest $L_{eff}$ is the highest (for $L= D_p$, $W=7$), it is evident that more detailed information about particle distributions must be incorporated for higher precision.  For $L = 0.25 D_p$, two values of $W \in \{6,8\}$ were tested, resulting in an increase in the $R^2$ score for higher $W$. However, the overall performance of the model with $L=0.25 D_p$ and $W= 8$ is only slightly higher than that of the model with $L=0.5 D_p$ and $W=4$. The model with $L=0.5 D_p$ and $W=5$ shows the best performance for drag and lift forces on all models tested and appears to be a good compromise between the accuracy and the computational costs needed to calculate the local volume fraction. The values of $L=0.5 D_p$ correspond to the maximal possible intersections of $k=27$ between a particle and neighboring \textit{NN cells}.

\begin{figure*}[t!]
    \centering
    \includegraphics{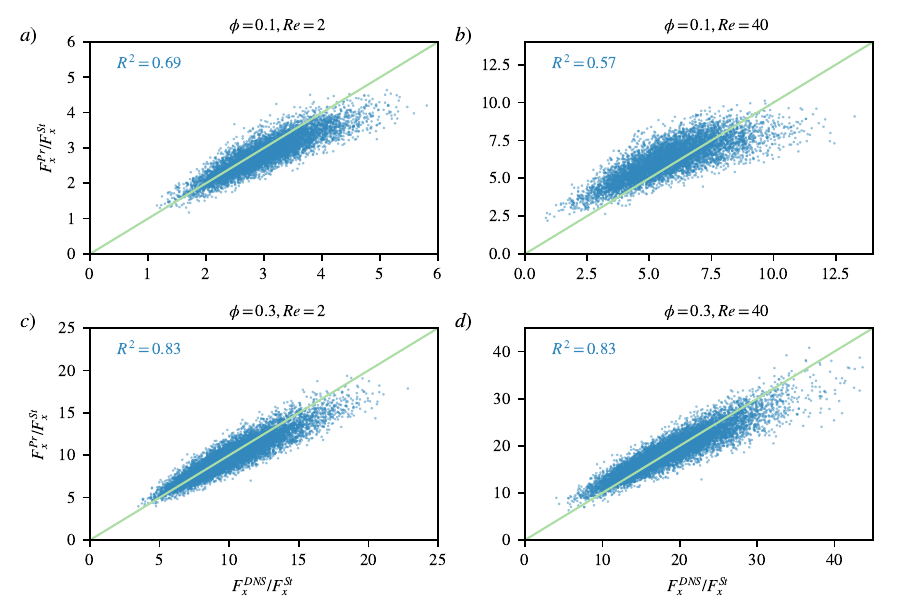}
    \caption{Comparison of DNS forces and forces obtained by the best \LVFB model with $L=0.5$ and $W=5$ for the limiting $Re$ and $\phi$ combinations. The $R^2$ scores for each correlation plot are indicated in the left top corner of each subfigure.}
    \label{fig:VF_0.5_5_correlation_plots}
\end{figure*}

\begin{figure}[t!]
    \centering
    \includegraphics[width=7cm]{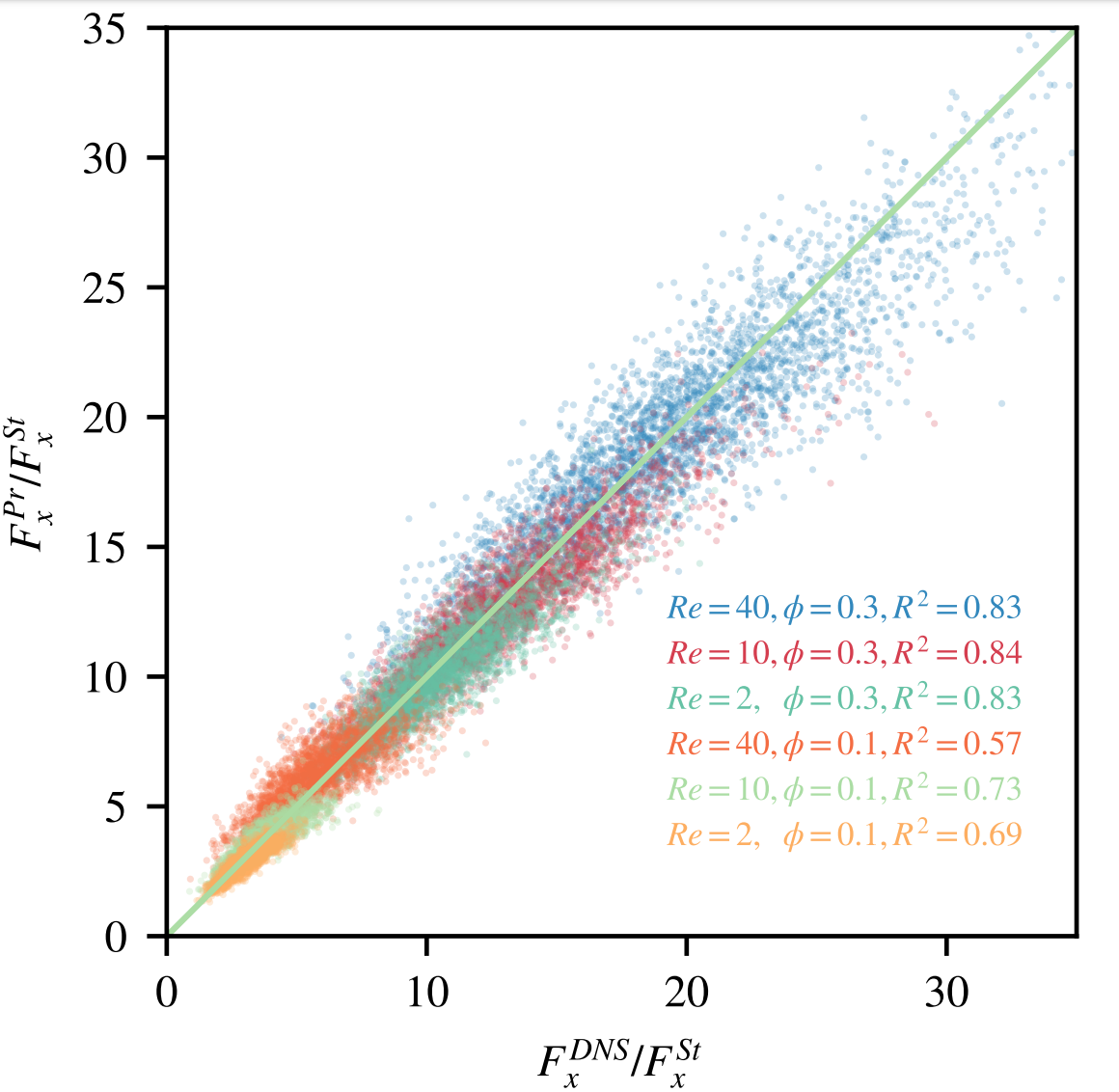}
    \caption{Comparison of DNS forces and forces obtained by the best \LVFB model with $L=0.5D_p$ and $W=5$ for $Re \in \{2,10,40\}$ and $\phi \in \{0.1,0.3\}$ combined in one plot. Each combination of $Re$ and $\phi$ corresponds to a specific color. Only one-third of the data points are presented for better visualization.}
    \label{fig:VF_0.5_5_combined_correlation_plot}
\end{figure}

Figure \ref{fig:VF_best_models_tables} shows the $R^2$ scores for the best models with \( L \in \{0.25 D_p, 0.5D_p, 1.0D_p\} \), evaluated in various combinations of $Re$ and $\phi$. Similarly to Figure \ref{fig:NP_best_models_tables}, a color-coded heat map is used to enhance the visualization of the corresponding $R^2$ scores. It is seen from Figure \ref{fig:VF_best_models_tables}, that the $R^2$ scores of a drag force for the individual combination of $Re$ and $\phi$ for the model with $L=D_p$ and $W=3$ vary between 0.13 and 0.63, indicating a model with low accuracy. Similar $R^2$ values are observed for the lift force, while the $R^2$ scores for the torque are even lower. The low accuracy for individual \( Re \)-\( \phi \) combinations, combined with the high \( R^2 \) score for the entire test data set, indicates that the model mainly captures the dependency of hydrodynamic forces on the Reynolds number ($Re$) and the volume fraction ($\phi$), while the influence of neighboring local particles on the forces is minimal. \review{Given the low accuracy of models with $L = D_p$, we did not include models with $L > D_p$ in our analysis, as they exhibit even lower accuracy.} On the other hand, models with $L \in \{0.25D_p, 0.5D_p\}$ show better accuracy for both force components as well as the lateral component of the torque. The $R^2$ scores of drag force for the model with $L=0.25D_p$ and $W=8$ range between 0.5 and 0.81, while for the lift force and torque similar precision values ($0.77<R^2<0.85$) are reported.  
The model with $L=0.5$ and $W=5$ shows the best accuracy, with $R^2$ scores for the drag force ranging between 0.57 and 0.85, while for the lift force the range is very similar and lays between 0.8 and 0.89. Interestingly, the three models presented in Figure \ref{fig:VF_best_models_tables} show a similar dependence of the model accuracy on the Reynolds number ($Re$). It is seen that with an increase in $Re$ the accuracy of the models decreases. This trend is clearly visible on the $R^2$ scores for the drag force $F_x$ for the model with $L=0.25D_p$, $W=8$, and values of $\phi = 0.1$. The same trend is observed for all values of $\phi$ for a model with $L=D_p$ and $W=3$. The dependency of the model accuracy on the volume fraction ($\phi$) is different for forces and torques. The drag force ($F_x$) exhibits a clear negative correlation between \( \phi \) and \( R^2 \) scores, whereas the torque \( T_l \) shows the opposite trend. In contrast, the results for the lift forces appear to be less sensitive to variations in \( \phi \), as no significant differences in \( R^2 \) scores in \( \phi \) have been observed for models with \( L \in \{0.25D_p, 0.5D_p\} \).

Each $R^2$ score visible in Figure \ref{fig:VF_best_models_tables} can also be illustrated by means of correlation plots, where the $x$-axis corresponds to the values of force/torque of the pr-DNS simulation from the test data set and the $y$-axis represents the same corresponding quantities predicted by the model. Figure \ref{fig:VF_0.5_5_correlation_plots} shows these correlation plots for the minimum and maximum values of $Re$ and $\phi$ for the drag force $F_x$ for the \LVFB model with $L=0.5D_p$ and $W=5$. The diagonal green line in each subplot represents the line of perfect agreement. For \(\phi=0.1\) and \(\text{Re}=2\) (Figure \ref{fig:VF_0.5_5_correlation_plots}a), the model shows a tendency to slightly underestimate the drag force ratio \(F^{\text{DNS}}_x/F^{\text{St}}_x\) for values greater than approximately 2.5, with most points lying below the green line in this range. In contrast, at the same $\phi=0.1$ but with $Re=40$, the model yields higher drag force values for \(F^{\text{DNS}}_x/F^{\text{St}}_x < 7\). The correlation for higher $\phi$, on the other hand, shows that the predicted forces are more symmetrical over the ideal green line, which is also indicated by higher values of the $R^2$ scores.

To provide a detailed explanation for the higher \( R^2 \) score observed for the entire test data set, as presented in Table \ref{tab:various_L_W_parameters}, we analyze the combined correlation plot for the data sets with \( Re \in \{2,10,40\} \) and \( \phi \in \{0.1,0.3\} \), shown in Figure \ref{fig:VF_0.5_5_combined_correlation_plot}. The plot demonstrates that the combined data points align more closely with the ideal green reference line, indicating improved agreement between the predicted and actual values of \( F^{DNS}_x \). Unlike the examples depicted in Figures \ref{fig:VF_0.5_5_correlation_plots}a and \ref{fig:VF_0.5_5_correlation_plots}b, the combined correlation plot does not exhibit systematic underprediction or overprediction in any specific range of \( F^{DNS}_x \). This alignment highlights the robustness of the predictive model in a wider range of flow conditions and particle volume fractions.

\begin{figure}[t!]
    \centering
    \includegraphics{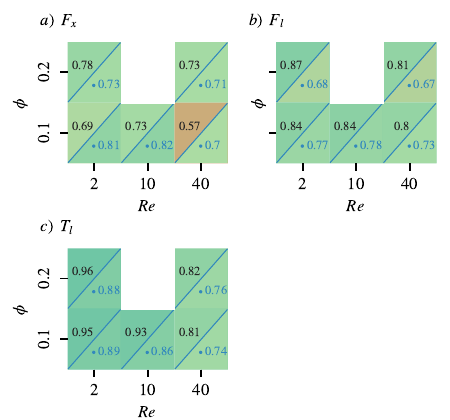}
    \caption{Comparison of $R^2$ scores between the \LVFB model with $L=0.5D_p$ and $W=5$ (upper left triangle of the table node) and the model described by \citet{seyed2021} (bottom right triangle of the table node).}
    \label{fig:VF_0.5_5_Seyed_compaarison}
\end{figure}

\begin{figure*}[!]
    \centering
    \includegraphics{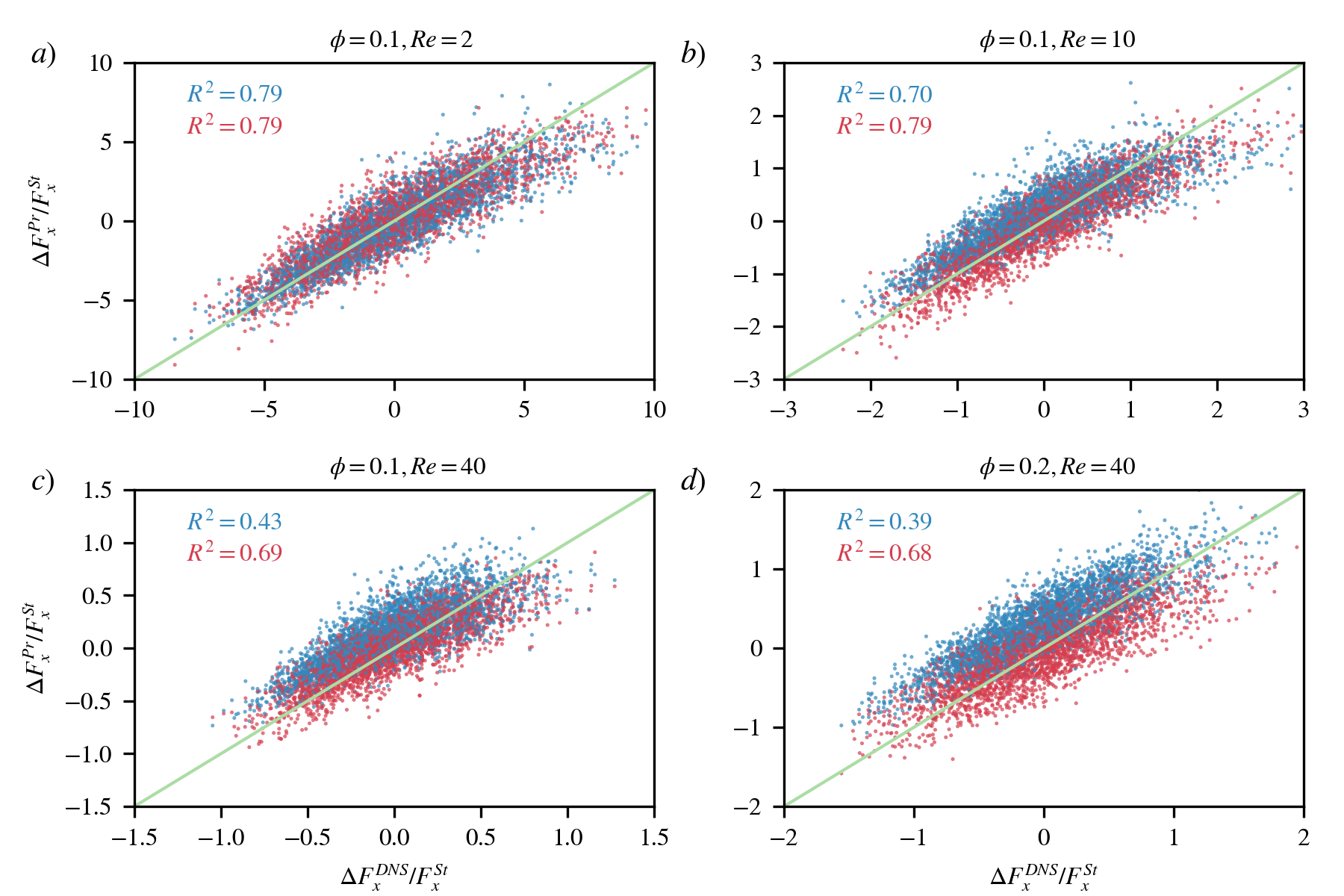}
    \caption{Comparison of regression plots for predictions from the present study's model (blue) and the model by \citet{seyed2021} (red) for common values of $Re$ and $\phi$. Each model was trained using its respective data set, while the test data set from \citet{seyed2021} was used for evaluation.}
    \label{fig:regression_plots_comparison}
\end{figure*}

\begin{figure*}[!]
    \centering
    \includegraphics{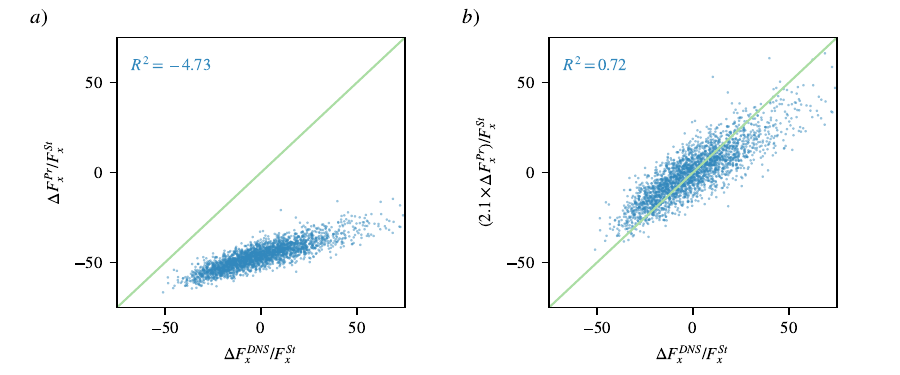}
    \caption{Correlation plot of the \LVFB model with $L=0.5$ and $W=5$ on the data set from \citet{seyed2021} for $Re=2$, $\phi=0.4$. Subfigure (a) shows predicted drag forces, while (b) presents the same data scaled by 2.1. The \LVFB model was trained on the data set from the present study.}
    \label{fig:Re_2_Phi_04_correlation}
\end{figure*}

\subsection{Comparison of LVFB models with existing models in the literature}
In order to compare the performance of the present model with those previously reported in the literature, we compare our best \LVFB model ($L=0.5D_P$ and $W=5$) with the results reported by \citet{seyed2021} in Figure \ref{fig:VF_0.5_5_Seyed_compaarison}. Note that each model was trained and evaluated on its own respective data set: our \LVFB model was trained and tested using the data set described in section \ref{sec:data_generation}, while the model from \citet{seyed2021} was trained and tested on the data set used in their study. The Figure presents the tables of $R^2$-scores for $F_x$, $F_l$, and $T_l$ across the range of $Re$ and $\phi$ parameters that overlap between the present work and the study by \citet{seyed2021}. Similarly to Figures \ref{fig:NP_best_models_tables} and \ref{fig:VF_best_models_tables}, the filling color of the table cell corresponds to the value of the $R^2$ score. Each cell in the figure is divided into two parts: the upper left shows the result of the present investigation, while the lower right represents the $R^2$ scores reported by \citet{seyed2021}.  For the drag force component ($F_x$), the model by \citet{seyed2021} consistently achieves higher $R^2$ scores for $\phi = 0.1$ in all Reynolds numbers, the difference being particularly noticeable at $Re = 40$, where our model shows a lower score of 0.57 compared to their 0.70. However, for $\phi = 0.2$, both models perform similarly well with only minor differences. For the lift force component ($F_l$), our \LVFB model shows slightly higher $R^2$ scores in all cases. The most noticeable difference is at $\phi = 0.2$, where our model achieves a $R^2$ score of 0.87 and 0.81 for $Re = 2$ and $Re=40$ compared to the 0.67 and 0.68 reported in \citet{seyed2021}. For the lateral torque component ($T_l$), our model consistently outperforms the model by \citet{seyed2021} under all conditions. In general, the comparison highlights that while the model by \citet{seyed2020} performs better for the predictions of the drag force at lower volume fractions, our model provides improved predictions for the components of the lift and torque throughout the parameter range.

To further evaluate the generalization and robustness of the \LVFB model (which we recall is trained on our test data set), we compare its predictions with the data set provided by \citet{seyed2021}. In that study, the authors include a publicly available repository that contains numerical data and codes to run the data-driven models that predict hydrodynamic forces and torques acting on particles. Using this resource, we tested the predictive capabilities of our model on the \citet{seyed2021} data set, ensuring a fair comparison between the two models. We used the \LVFB model with $L = 0.5D_P$ and $W=5$ and compared the drag forces predicted by our model and the model of \citet{seyed2021}. The resulting correlation plots of the drag force of our model and the model of \citet{seyed2021} are presented in Figure \ref{fig:regression_plots_comparison} for the following parameters: $Re \in \{2, 10, 40\}$ and $\phi \in \{0.1, 0.2\}$. Note that, unlike in Figures \ref{fig:VF_0.5_5_correlation_plots} and \ref{fig:VF_0.5_5_combined_correlation_plot}, we now subtract the mean pr-DNS drag force from both the predicted model forces and pr-DNS to align our analysis with the approach of \citet{seyed2021}. As a result, we express the forces as \(\Delta F_x^{Pr} = F_x^{Pr} - \langle F_x^{DNS} \rangle\) and \(\Delta F_x^{DNS} = F_x^{DNS} - \langle F_x^{DNS} \rangle\). Since the same constant (\(\langle F_x^{DNS} \rangle\)) is subtracted from both parameters, this adjustment does not affect the resulting \(R^2\) scores. It is seen from Figure \ref{fig:regression_plots_comparison}a, that for the lowest values of $Re$ and $\phi$, our model produces correlation plots almost identical to those of \citet{seyed2020}, with matching $R^2$ scores.  However, as the Reynolds number increases, the deviations between the two models become more pronounced. At \( Re = 10 \), the difference remains relatively small, but for \( Re = 40 \) in both volume fractions, our model exhibits a significantly lower \( R^2 \) score than the model of \citet{seyed2021}. We attribute this deviation primarily to the differences in the mean drag force of the training data sets. It is seen from Figure \ref{fig:validation}, that while the difference in mean drag force is minimal for $Re=2$ and $\phi=0.1$, it becomes more pronounced with higher $Re$, which is even more amplified for $\phi = 0.2$. This is also supported by the fact that, despite the upward shift in the \LVFB model's predictions, the data follow a similar trend and remain well aligned with the perfect correlation line (green), indicating that the model accurately captures the overall force distribution. These results also confirm that our model does not suffer from overfitting, as it maintains predictive accuracy even when applied to external test data, as shown in Figure \ref{fig:regression_plots_comparison}.

\begin{figure*}[t!]
    \centering
    \includegraphics{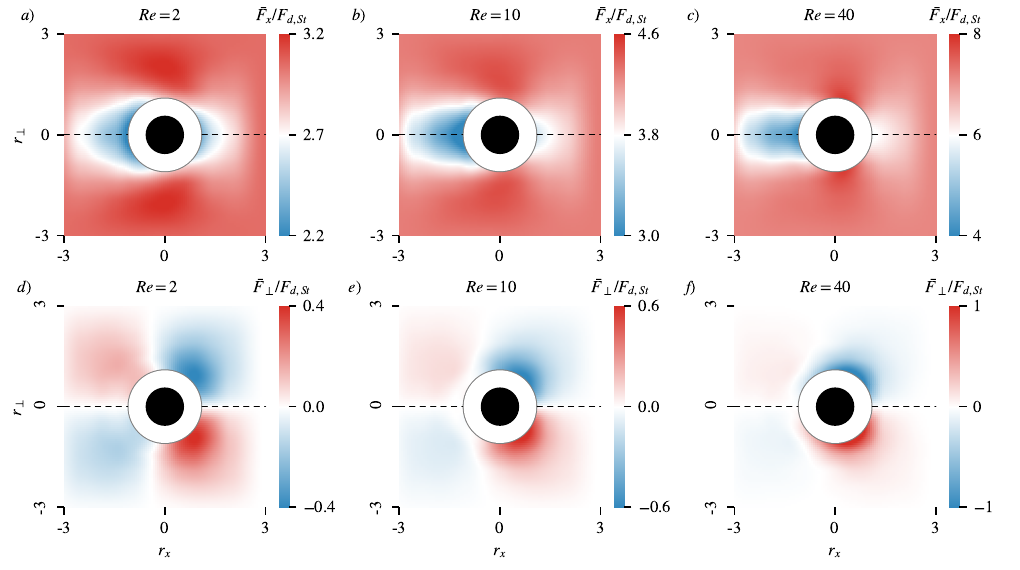}
    \put(-485,230){\vector(1,0){15}}
    \put(-485,235){$u_{\infty}$}
    \caption{The predicted $\bar{F}_x$ ($a$, $b$, $c$) and $\bar{F}_\perp$ ($d$, $e$, $f$) 
    forces acting on the reference particle with the presence of one neighboring particle located at $(r_x,r_\perp)^T$. The reference particles is located at $r_x = 0$, $r_\perp = 0$ and shown as black filled circle. The gray circle indicates the closest possible position of the center of the neighboring particle. The flow direction is similar for all subplots and indicated by the vector $u_\infty$ next to subfigure $a)$.}
    \label{fig:Fx_Fy_neighbouring_part}
\end{figure*}

Beyond evaluating our model within the parameter space tested, we also explored its ability to extrapolate predictions to volume fractions outside the training range. Specifically, we evaluated its performance at \( Re = 2 \) and \( \phi = 0.4 \) using simulation data from \citet{seyed2021}. Figure \ref{fig:Re_2_Phi_04_correlation}a compares the drag forces (\( F_x^{Pr} \)) predicted by our model with the corresponding simulation data. It is visible that the \LVFB model tends to systematically underestimate the drag force under these conditions, resulting in a poor \( R^2 \) score of -4.73. This was expected, since we are in range of extrapolation for our parameter space explored thus far. However, the observed shift suggests that applying a scaling correction could improve the performance. Using systematically different scaling factors, we determined that multiplying the model predictions by a factor of 2.1 significantly improves the agreement, as shown in Figure \ref{fig:Re_2_Phi_04_correlation}b. This correction leads to a \( R^2 \) score comparable to the values observed for \( Re = 2 \) within the training range. These findings suggest that, while the model retains generalizability to new conditions, systematic biases can emerge when extrapolating beyond the training set. However, the fact that a simple scaling adjustment substantially improves accuracy indicates that the underlying physical relationships are well captured. 

\subsection{Interpretability of the \LVFB model}

To analyze the pairwise interaction between the reference particle and a single neighboring particle, we construct two-dimensional force maps where the axes represent the relative position of a neighboring particle with respect to the reference particle. These relative positions are defined as \( r_x \) and \( r_{\perp} \), corresponding to the streamwise and lateral components, respectively. The streamwise component is aligned with the \( x \)-axis and defined as \(r_x = (x_p^n - x_p^r)\), where $x_p^n$ is the $x$ position of the neighboring particle and $x_p^r$ is the $x$ position of the reference particle. The lateral component, \( r_{\perp} \), is perpendicular to \( r_x \) and can be oriented differently. To specify the lateral direction, we introduce the angle \(\theta\), where \( \theta = 0 \) aligns \( r_{\perp} \) with the \( y \)-axis and \( \theta = \pi/2 \) aligns it with the \( z \)-axis. Hence, for \( \theta = 0 \) the lateral component of the relative position is calculated as \( r_{\perp} = (y_p^n - y_p^r) \), where $y_p^n$ is $y$ position of the neighboring particle and $y_p^r$ is $y$ position of reference particle. Similarly, \( r_{\perp} = (z_p^n - z_p^r) \) for \( \theta = \pi/2 \). The lateral component of the force $F_\perp$, similarly to $r_\perp$, also depends on the angle $\theta$ and represents the force component perpendicular to the axis $x$. We select four values, \( \theta \in \{0, \pi/2, \pi, 3\pi/2\} \), and construct separate force maps for the drag, \( F_x(r_x, r_{\perp}) \), and the lateral force, \( F_{\perp}(r_x, r_{\perp}) \). Subsequently, these force maps are averaged, resulting in:
\begin{equation}
    \bar{F}_x(r_x, r_{\perp}) = \frac{1}{4} \sum_{\theta_i} F^i_x(r_x, r_{\perp}),
\end{equation}
\begin{equation}
    \bar{F}_{\perp}(r_x, r_{\perp}) = \frac{1}{4} \sum_{\theta_i} F^i_{\perp}(r_x, r_{\perp}), 
\end{equation}

where the index $i$ corresponds to different angles of $\theta$. This can be referred to as azimuthal averaging (which is used by \citet{seyed2021} and adopted by \citet{cheng2023physics}) with an increment step of $\pi/2$. Figure \ref{fig:Fx_Fy_neighbouring_part} ($a$, $b$, $c$) shows that if the neighboring particle is located in front of the reference particle (i.e. $r_x < -1$ and $
|r_\perp| < 1$), the drag force ($\bar{F_x}$) experienced by the reference particle is significantly reduced. On the other hand, if the neighboring particle is located within the same $r_x$ coordinate and with $1 < r_\perp < 3$, the reference particle experiences an amplified drag force, which is observed for all Reynolds numbers. We also observe similar patterns in other studies \cite{balachandar2020,seyed2021,cheng2023physics}. Regarding the lift force ($\bar{F_\perp}$) experienced by the reference particle, the subplots ($d$,$e$,$f$) in Figure \ref{fig:Fx_Fy_neighbouring_part} are also consistent with the similar figure of \citet{cheng2023physics} in all $Re$. It is important to note that the axisymmetric properties of the model are not explicitly set, but appear as the result of averaging of several planes, which coincide with streamwise direction $x$ and intersect the center of reference particle. The non-averaged contour plots do not show perfect symmetrical distribution neither over $xy$-nor over $xz$-planes. As was mentioned in \cite{seyed2021} and later confirmed by \cite{cheng2023physics}, in order to achieve a perfect axisymmetry, one needs to perform rotation data augmentation, within which particle positions are transformed by rotation over streamwise direction. In the present study, we perform data augumentation only by translating the position of particles. This could explain, why we did not reach the perfect symmetry within individual planes. 

\subsection{\review{Computational efficiency of the LVBF model}}
\review{To evaluate the computational efficiency of the LVFB model, we performed a strong scaling analysis and compared its execution time with that of a point-particle simulation. For this analysis, we used 2 nodes from the Barnard HPC cluster provided by the NHR Center of TU Dresden\cite{ZIH2025}. Each computation node consists of two Intel Xeon Platinum 8470 CPUs, with a total of 104 cores per node. To perform the point-particle simulation, we used the same computational scenario and numerical code as reported by \citet{zhao2021flocculation}. The simulation setup includes $N_p=10\ 000$ randomly distributed point particles with diameter $D_p$ placed in the domain of $L_x \times L_y \times L_z = (125D_p)^3$. The discretization $N_x \times N_y \times N_z = (128)^3$ of the numerical domain is uniform and isotropic, which results in the size of a computational cell to be approximately equal to the particle diameter. The domain has triple periodic boundary conditions, and the flow is controlled by an additional forcing term that induces isotropic turbulence. We refer the reader to \citet{zhao2021flocculation} for the exact definition of the simulation case (labeled as \textit{"Flo5"} in the same publication) and the numerical methods used for the point-particle simulation approach.}

\begin{figure}[b!]
    \centering
    \includegraphics{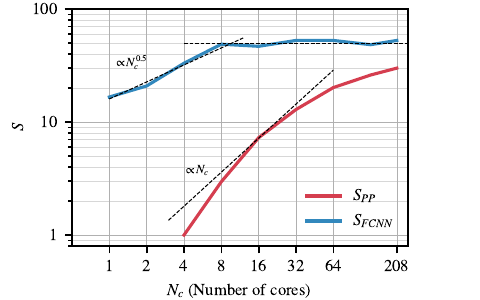}
    \caption{Speedup $S$ as a function of number of cores of point-particle simulation ($S_{PP}$) and \LVFB model ($S_{FCNN}$).}
    \label{fig:LVFB_performance}
\end{figure}
\review{
Within the strong scaling, we compare the execution time of a single time step of the point-particle simulation ($\Delta t_{it}^{PP}$) for different numbers of cores ($N_c$) used for parallelization, i.e., $\Delta t_{it}^{PP} = f(N_c)$. We express the execution time in terms of speedup $S$, which is calculated for point-particle simulation as
\begin{equation}
    S_{PP}(N_c) = \frac{\Delta t_{it}^{PP}(4)}{\Delta t_{it}^{PP}(N_c)},
\end{equation}
where $\Delta t_{it}^{PP}(4)$ is the execution time of one iteration using four cores, which we use as the reference time scale. Figure \ref{fig:LVFB_performance} shows $S_{PP}$ for the number of cores in the range between 4 and 208.
Note that speedup is inversely proportional to $\Delta t_{it}^{PP}(N_c)$, which means that higher values of $S$ correspond to better performance. It is seen from Figure \ref{fig:LVFB_performance} that $S_{PP}$ between 4 and 16 processors increases faster than ideal scaling (that is, $S \propto N_c$).
This is due to the domain decomposition in our parallelized application, where an increasing number of cores reduces the number of possible neighboring particles in a subdomain. Note that in our application, the number of subdomains is equal to the number of cores used. For the small number of subdomains, the simulation scales to $\propto N_p^2$, making the identification of neighbors a performance bottleneck. However, with a high number of subdomains, the computational cost of the Lagrangian operations scales similarly to that of the Eulerian operations, i.e. approximately as $\propto N_p$. As the communication overhead between parallel tasks increases with the number of cores, the speedup begins to fall below the ideal linear scaling (i.e. $\propto N_c$) starting from $N_c > 16$ and continues to degrade with increasing $N_c$.}

\review{We also compute the speedup $S$ for the best \LVFB  model (e.g. with $L=0.5$ and $W=5$) which we define with  the same reference time $\Delta t_{it}^{PP}(4)$:
\begin{equation}
    S_{FCNN} = \frac{\Delta t_{it}^{PP}(4)}{\Delta t_{it}^{FCNN}(N_c)},
\end{equation}
where $\Delta t_{it}^{FCNN}$ is the time needed for the \LVFB model to predict the hydrodynamic forces and torques for $10\ 000$ particles. It is seen from Figure \ref{fig:LVFB_performance} that the speedup of the \LVFB model scales $S\propto \sqrt{N_c}$ in the range between 1 and 8 cores. For $N_c > 8$ the \LVFB model does not show the increase in $S$ with further parallelization. Despite the lower speed-up scaling values, we obtain a scaling, where $S_{FCNN}$ remains above the $S_{PP}$ curve. This means that the absolute values of $S_{FCNN}$ are higher for the entire range of $N_c$. Hence, the \LVFB model requires less execution time than the time required by the point-particle simulation. For $N_c = 4$  the ratio $\Delta t_{it}^{FCNN}/\Delta t_{it}^{PP} = 0.03$, while for $N_c = 208$ the same ratio is equal to $0.57$.}

\review{This suggests that for domain sizes on the order of $10^2 D_p$ and particle counts on the order of $10^4$, the peak computational cost of the \LVFB model becomes comparable to that of conventional point-particle simulations. The analysis highlights that the cost of the FCNN-based model remains within a reasonable scale and is not prohibitively high. Nonetheless, it should be acknowledged that the parallel scaling of the FCNN model is far from optimal, and future work should aim at improving its computational efficiency. That said, because the model inputs are independent, predictions for different particles can also be computed in parallel across multiple jobs, which offers an additional layer of parallelism. It is important to note, however, that the primary focus of the present study was not on performance optimization but rather on achieving high predictive accuracy of the model.}

\review{}
\section{Conclusion}
\label{sec:conclusion}
The advancement of numerical methods for predicting various types of flow plays a crucial role in understanding microphysics and optimizing engineering structures and devices. Recent studies indicate that integrating machine learning techniques into the prediction of hydrodynamic forces on particles in multiphase flows can significantly enhance the accuracy of conventional computational fluid dynamics methods.  

In this study, we conducted multiple sets of pr-DNS simulations and trained two types of FCNN models: the \PPB models, which directly incorporate the locations of neighboring particles in the input layer, and the \LVFB models, which utilize local volume fractions computed on an additional NN grid, a structured rectangular mesh with a resolution comparable to the diameter of the particle. First, we demonstrate that by using a larger data set in conjunction with sophisticated techniques such as dropout, batch normalization, and cyclic learning rate, it is possible to achieve significantly higher accuracy than previously reported for FCNNs. Second, we proposed a novel input layer construction approach based on local volume fraction values. We showed that for a similar neural network architecture, this approach yields substantially higher \( R^2 \) scores than previously tested \PPB models. Furthermore, we validated the \LVFB model on an external data set and found that it achieved similar \( R^2 \) scores when the mean drag forces closely matched our training data. However, in cases where the mean drag force differed more significantly, the \( R^2 \) scores were lower, although the general trend was captured correctly. This discrepancy was attributed to a shift in the predicted values, likely due to the higher average drag forces in our training data set.

\review{It is also important to highlight that the current FCNN architecture was chosen to enable a direct comparison between the \PPB and \LVFB approaches. Although we observe no signs of overfitting, more advanced architectures, such as those incorporating convolutional layers or physics-informed networks (as used by \citet{seyed2021,cheng2023physics}), may significantly improve accuracy. CNNs, in particular, are well-suited for structured input like the \LVFB grid and scale better with higher resolutions. Future work could also explore architectures such as attention-based models (e.g., Transformers) or Mixture-of-Experts (MoE) networks to adaptively capture flow-regime-specific features.}

Overall, this new input vector construction method may be particularly beneficial for modeling more complex systems, such as those involving polydisperse particles of various shapes. Unlike \PPB approaches, this method does not require special treatment to account for particle shape and size, as these characteristics are inherently captured through the representation of the local volume fraction. 
\review{Toward this end, only small modifications are needed to adjust \LVFB model for polydispersity and particle shape variability: The input vector should be augmented with properties of the particle of interest (i.e., the one for which force is being predicted), such as its size, shape, and orientation. This allows the model to condition its prediction on both the particle-specific attributes and the local environment.} Furthermore, this approach could be advantageous when incorporating wall boundary conditions, as external walls can also be effectively represented through local volume fractions. \review{We believe that developing a modified \LVFB model that considers the effects of polydispersity and the wall boundary and testing it in comparison with pr-DNS and point-particle modeling is a promising direction for future work.}

\begin{acknowledgments}
The authors gratefully acknowledge the support through the German Research Foundation (DFG) grant VO2413/2-1 and the Agnes-Pockels Fellowship of TU Braunschweig. The work used the supercomputer Phoenix and was supported by the Gauß-IT-Zentrum of the University of Braunschweig (GITZ). The authors gratefully acknowledge the Gauss Centre for Supercomputing e.V. (www.gauss-centre.eu) for funding this project by providing computing time on the GCS Supercomputer SUPERMUC-NG at Leibniz Supercomputing Centre (www.lrz.de). 
The authors gratefully acknowledge the computing time made available to them on the high-performance computer at the NHR Center of TU Dresden. This center is jointly supported by the Federal Ministry of Education and Research and the state governments participating in the NHR (www.nhr-verein.de/unsere-partner).
\end{acknowledgments}

\section*{Data Availability Statement}
The train and test data set described in Section \ref{sec:data_generation}, the best trained models that are shown in Sections \ref{sec:ppb_performance} and \ref{sec:lvfb_performance}, and the Python scripts which allow one to test different models within different data sets are included in the open source repository \url{https://github.com/metialex/HydroForces-ML/}

\bibliography{manuscript}
\end{document}